\documentclass[12pt]{article}
 \usepackage[cp1251]{inputenc} 
 \usepackage[english]{babel}
 \usepackage{amsmath, latexsym, amssymb, bm, array, graphics, eucal,
 amsfonts,}
 \makeatletter
 \renewcommand{\@biblabel}[1]{#1.\hfill}
 \renewcommand{\Re}{\mathop{\rm Re}}
 \renewcommand{\Im}{\mathop{\rm Im\,}}
\renewcommand{\div}{\mathop{\rm div}}
\newcommand{\mc}[1]{\mathcal{#1}}
\newcommand{\E}{\mc{E}}
 \makeatother
 \renewcommand{\baselinestretch}{1.22}
\usepackage[dvips]{graphicx}

\begin{document}
 \thispagestyle{empty}
 \renewcommand{\abstractname}{\ }
 \large
 \renewcommand{\refname}{\begin{center} REFERENCES\end{center}}
\newcommand{\const}{\mathop{\rm const\, }}

 \begin{center}
\bf TRANSVERSE ELECTRIC CONDUCTIVITY OF QUANTUM COLLISIONAL PLASMAS
\end{center}\medskip
\begin{center}
  \bf  A. V. Latyshev\footnote{$avlatyshev@mail.ru$},
  A. A. Yushkanov\footnote{$yushkanov@inbox.ru$}
\end{center}\medskip

\begin{center}
{\it Faculty of Physics and Mathematics,
\\ Moscow State Regional University,  105005,\\ Moscow, Radio str., 10--A}
\end{center}\medskip

\begin{abstract}
Formulas for calculation of transverse dielectric function and
transverse electric conductivity in quantum collisional plasmas under
arbitrary degree of dege\-ne\-racy of the electron gas are received.
The Wigner -- Vlasov -- Boltzmann kinetic equation with collision
integral in BGK (Bhatnagar, Gross and Krook) form
in coordinate space is used. Various special cases are
in\-ves\-ti\-gated. The case of
fully degenerate quantum plasma was considered separately.
Comparison with Lindhard's formula has been realized. \medskip

{\bf Key words:} collisional plazma, Schr\"{o}dinger equation,
electric conductivity, dielectric function, Lindhard
formula.\medskip

PACS numbers:  52.25.Dg Plasma kinetic equations,
52.25.-b Plasma properties, 05.30 Fk Fermion systems and
electron gas.

\end{abstract}

\begin{center}\bf
  1. INTRODUCTION
\end{center}

In the present work formulas for calculation of electric
conductivity and dielectric function in quantum collisional
plasma under arbitrary tempe\-ra\-tu\-re, i.e. under arbitrary degree of
degeneration of the electron gas are deduced.

During the derivation of the kinetic equation we generalize the
approach, developed by Klimontovich and Silin \cite {Klim}.

Dielectric function in the collisionless quantum gaseous plasma
was studied by many authors (see, for example, \cite {Klim} --
\cite{Shukla2}).

In the work \cite {Manf}, where the one-dimensional case
of the quantum plasma is investigated, the importance of derivation
of dielectric function with use of the quantum kinetic equation
with collision integral in the form of BGK -- model (Bhatnagar,
Gross, Krook) \cite{BGK}, \cite{Opher} was noted.

The present work is devoted to the performance of this problem.

A dielectric function is one of the most significant
characteristics of a plasma. This quantity is necessary for
description of the skin effect \cite{Gelder},
for analysis of surface plasmons \cite{Fuchs},
for description of the process of propagation and damping of the
transverse plasma oscillations \cite{Shukla2},
the mechanism of electromagnetic
waves penetration in plasma \cite{Shukla1}, and
for analysis of other problems of plasma physics \cite{Fuchs2},
\cite{Dressel}, \cite{Wier}, \cite{Brod} and \cite{Manf2}.

Kliewer and Fuchs were the first who have noticed \cite {Kliewer},
that the dielectric function for quantum plasma deduced by Lindhard
in collisional case does not pass into dielectric function for
classical plasma in the limit when Planck's constant $\hbar$
converges to zero. This means, that dielectric Lindhard's function
does not take into account electron collisions correctly. Kliewer
and Fuchs have corrected Lindhard's dielectric function "by hands" \,
so that it passed into classical one under condition $ \hbar \to 0$.

In the works  \cite{Fuchs}, \cite{Fuchs2} the dielectric function
received by them was applied to consideration of various questions
of  metal optics.

In the work \cite{Mermin} the correct account of collisions in
framework of the relaxation model in electron momentum space
for the case of longi\-tu\-di\-nal dielectric function has been carried out. At
the same time the correct account of influence of collisions for
transverse dielectric function has not been implemented till
now.

The aim of the present work is the elimination of this lacuna. 

\begin{center}
  {\bf 2. KINETIC EQUATION FOR THE WIGNER FUNCTION}
\end{center}

Let's consider the Schr\"{o}dinger equation written for a particle
in an elect\-ro\-magnetic field in terms of density matrix $\rho$
$$
i\hbar \dfrac{\partial \rho}{\partial t}=H\rho-{H^*}'\rho.
\eqno{(1.1)}
$$

Here $H$ is the Hamilton operator, $H^*$ is the complex conjugate
operator to $H$, ${H^*}'$ is the complex conjugate operator to the
$H$, which forces on primed spatial variables $\mathbf{r}'$.

Hamilton operator for the free particle which is in the field of the
scalar potential $U$ and in the field of vector potential
$\mathbf{A}$, has the following form
$$
H=\dfrac{(\mathbf{p}-\dfrac{e}{c}\mathbf{A})^2}{2m}+eU=$$$$=
\dfrac{\mathbf{p}^2}{2m}-
\dfrac{e}{2mc}(\mathbf{p}\mathbf{A}+\mathbf{A}\mathbf{p})
+\dfrac{e^2}{2mc^2}\mathbf{A}^2+eU.
\eqno{(1.2)}
$$

Here $\mathbf{p}$ is the momentum operator, $\mathbf{p}=-i\hbar \nabla$,
$e$ is the electron charge, $m$ is the electron mass, $c$ is the
light velocity.

Let's rewrite the Hamilton operator (1.2) in the explicit form
$$
H=-\dfrac{\hbar^2}{2m}\triangle+\dfrac{ie\hbar}{2mc}\Big(2\mathbf{A}\nabla
+\nabla\mathbf{A}\Big)+
\dfrac{e^2}{2mc^2}\mathbf{A}^2+eU.
\eqno{(1.3)}
$$
\smallskip

Complex conjugate to the $H$ operator $H^*$ according to (1.3) has
the form
$$
H^*=-\dfrac{\hbar^2}{2m}\triangle-\dfrac{ie\hbar}{2mc}
\Big(2\mathbf{A}\nabla+\nabla\mathbf{A}\Big)+
\dfrac{e^2}{2mc^2}\mathbf{A}^2+eU.
$$
\smallskip

Hence we can write down  for $H\rho$
$$
H\rho=-\dfrac{\hbar^2}{2m}\Delta \rho+\dfrac{ie\hbar}{2mc}
\Big(2\mathbf{A}\nabla \rho+\rho\nabla\mathbf{A}\Big)+
\dfrac{e^2}{2mc^2}\mathbf{A}^2\rho+eU \rho
\eqno{(1.4)}
$$\smallskip
and for ${H^*}'\rho$
$$
{H^*}'\rho=-\dfrac{\hbar^2}{2m}\Delta'\rho-\dfrac{ie\hbar}{2mc}
\Big(2\mathbf{A'}\nabla'\rho+\rho\nabla'\mathbf{A}\Big)+
\dfrac{e^2}{2mc^2}\mathbf{A'}^2\rho+eU' \rho.
\eqno{(1.5)}
$$

Operators $\nabla$ and $\Delta$ from Eqs (1.4) and (1.5) force on
unprimed spatial variables of the density matrix, i.e.
$\nabla=\nabla_{\mathbf{R}}$, $\Delta=\Delta_{\mathbf{R}}$.
In the operator ${H^*}'$ is necessary to replace the operators
$\nabla=\nabla_{\mathbf{R}}$ and $\Delta=\Delta_{\mathbf{R}}$ by
operators $\nabla'\equiv\nabla_{\mathbf{R}'}$ and
$\Delta'\equiv\Delta_{\mathbf{R}'}$, in addition we introduce the
following designations
$$
\mathbf{A'}\equiv \mathbf{A}(\mathbf{R'},t),
\quad
\quad U'\equiv U(\mathbf{R'},t).
$$

Let's find the right-hand member of the equation (1.1), i.e.
difference between relations (1.4) and (1.5): $H\rho-{H^*}'\rho$.
According to (1.4) and (1.5) we have
$$
H\rho-{H^*}'\rho=-\dfrac{\hbar}{2m}\Big(\Delta \rho-\Delta'\rho\Big)+
\hspace{8cm}
$$\smallskip
$$
\hspace{1cm}+\dfrac{i e \hbar}{2mc}
\Big[2\Big(\mathbf{A}\nabla \rho+
\mathbf{A'}\nabla'\rho\Big)+\rho\Big(\nabla \mathbf{A}+
\nabla'\mathbf{A}\Big)\Big]+
$$\smallskip
$$
\hspace{2cm}
+\dfrac{e^2}{2mc^2}\Big[\mathbf{A}^2(\mathbf{R},t)-\mathbf{A}^2(
\mathbf{R'},t)\Big]+
e[U(\mathbf{R},t)-U(\mathbf{R'},t)]\rho.
$$
\smallskip

The connection between density matrix
$\rho(\mathbf{r},\mathbf{r}',t)$ and Wigner function
\cite{Wigner}, \cite{Tatarskii}, \cite{Hillery}
$f(\mathbf{r},\mathbf{p},t)$ is
given by the inversion and direct  Fourier conversions
$$
f(\mathbf{r},\mathbf{p},t)=\int
\rho(\mathbf{r}+\dfrac{\mathbf{a}}{2},\mathbf{r}-
\dfrac{\mathbf{a}}{2},t)e^{-i\mathbf{p}\mathbf{a}/\hbar}d^3a,
$$\smallskip
$$
\rho(\mathbf{R},\mathbf{R}',t)=\dfrac{1}{(2\pi \hbar)^3}
\int f\Big(\dfrac{\mathbf{R}+\mathbf{R}'}{2}, \mathbf{p},t\Big)
e^{i\mathbf{p}(\mathbf{R}-\mathbf{R}')/\hbar}d^3p.
$$
\smallskip

The Wigner function is analogue of distribution function for quantum
systems. It is widely used in the diversified  physics
questions. Wigner's function was investigated, for example,
in works \cite {Arnold} and \cite {Kozlov}.

Substituting the representation of the density matrix in terms
of the Wigner function (1.2) into the equation for the density matrix
(1.1), we obtain
$$
i\hbar\dfrac{\partial \rho}{\partial t}=H\left\{\dfrac{1}{(2\pi\hbar)^3}
\int f\Big(\dfrac{\mathbf{R}+\mathbf{R}'}{2},\mathbf{p'},t\Big)
e^{i\mathbf{p'}(\mathbf{R}-\mathbf{R}')/\hbar}\,d^3p'\right\}-
$$
$$
\hspace{1.8cm}-{H^*}'\left\{\dfrac{1}{(2\pi\hbar)^3}
\int f\Big(\dfrac{\mathbf{R}+\mathbf{R}'}{2},
\mathbf{p'},t\Big)
e^{i\mathbf{p'}(\mathbf{R}-\mathbf{R}')/\hbar}\,d^3p'\right\}.
$$

Let's use the equalities written above. Thus the right--hand member
of the previous equation we may present in explicit form.  As a
result we receive the following equation
$$
i\hbar \dfrac{\partial \rho}{\partial t}=
\dfrac{1}{(2\pi \hbar)^3}\int\left\{-\dfrac{i\hbar}{m}
\mathbf{p'}\nabla f +
\dfrac{ie\hbar}{2mc} \Big[\div{\mathbf{A}(\mathbf{R},t)}+
\div{\mathbf{A}(\mathbf{R}',t)}\Big]f+
\right.
$$\smallskip
$$
\hspace{1cm}+\left.\dfrac{ie\hbar}{2mc}\Big[\mathbf{A}(\mathbf{R},t)+
\mathbf{A}(\mathbf{R}',t)\Big]
\nabla f-\dfrac{e}{mc}\Big[\mathbf{A}(\mathbf{R},t)-
\mathbf{A}(\mathbf{R}',t)\Big]
\mathbf{p'}f\right.+
$$\smallskip
$$
+\left.\dfrac{e^2}{2mc^2}\big[\mathbf{A}^2(\mathbf{R},t)-
\mathbf{A}^2(\mathbf{R}',t)\big]f+\right.
$$\smallskip
$$
+\left.e\big[U(\mathbf{R},t)-U(\mathbf{R}',t)\big]f\right\}
e^{i\mathbf{p'}(\mathbf{R}-\mathbf{R}')/\hbar}
d^3p'.
\eqno{(1.6)}
$$
\smallskip

In the equation (1.6) we will put
$$
\mathbf{R}=\mathbf{r}+\dfrac{\mathbf{a}}{2},\qquad
\mathbf{R}'=\mathbf{r}-\dfrac{\mathbf{a}}{2}.
$$

Then in this equation we obtain
$$
 f\Big(\dfrac{\mathbf{R}+\mathbf{R}'}{2}, \mathbf{p'},t\Big)
e^{i\mathbf{p'}(\mathbf{R}-\mathbf{R}')/\hbar}=f(\mathbf{r}, \mathbf{p'},t)
e^{i\mathbf{p'\,a}/\hbar}.
$$

Let's multiply the equation (1.6) by
$e^{-i\mathbf{p}\mathbf{a}/\hbar}$ and let's integrate it by
$\mathbf{a}$. Then we will divide both parts of the equation by
$i\hbar$. As a result we receive
$$
\dfrac{\partial f}{\partial t}=
\iint\Bigg\{-\dfrac{\mathbf{p'}}{m}\nabla f +
\dfrac{e}{2mc}
\Big[\mathbf{A}(\mathbf{r}+\dfrac{\mathbf{a}}{2},t)+
\mathbf{A}(\mathbf{r}-\dfrac{\mathbf{a}}{2},t)\Big]\nabla
f+
$$\smallskip
$$
+\dfrac{ie}{mc\hbar}\Big[
\mathbf{A}(\mathbf{r}+\dfrac{\mathbf{a}}{2},t)
-\mathbf{A}(\mathbf{r}-\dfrac{\mathbf{a}}{2},t)\Big]\mathbf{p'} f+
$$\smallskip
$$+
\dfrac{e}{2mc}\Big[\div{\mathbf{A}(\mathbf{r}}+\dfrac{\mathbf{a}}{2},t)+
\div{\mathbf{A}(\mathbf{r}-\dfrac{\mathbf{a}}{2},t)}\Big]f-
$$\smallskip
$$
-\dfrac{ie^2}{2mc^2\hbar}\Big[\mathbf{A}^2(\mathbf{r}+
\dfrac{\mathbf{a}}{2},t)-
\mathbf{A}^2(\mathbf{r}-\dfrac{\mathbf{a}}{2},t)\Big]f-
$$\smallskip
$$
-\dfrac{ie}{\hbar}\Big[U(\mathbf{r}+\dfrac{\mathbf{a}}{2},t)-
U(\mathbf{r}-\dfrac{\mathbf{a}}{2},t)\Big]f\Bigg\}
e^{i(\mathbf{p'}-\mathbf{p})\mathbf{a}/\hbar}
\dfrac{d^3a\,d^3p'}{(2\pi\hbar)^3}.
\eqno{(1.7)}
$$
\smallskip

On the left-hand side of the equation (1.7) we have
$f=f(\mathbf{r},\mathbf{p},t)$, in the integral we have
$f=f(\mathbf{r},\mathbf {p'},t)$.

We consider the integral
$$
\iint\mathbf{p'}(\nabla f)e^{i(\mathbf{p'}-\mathbf{p})\mathbf{a}/\hbar}
\dfrac{d^3a\,d^3p'}{(2\pi\hbar)^3}=
\nabla\iint\mathbf{p'} f e^{i(\mathbf{p'}-\mathbf{p})\mathbf{a}/\hbar}
\dfrac{d^3a\,d^3p'}{(2\pi\hbar)^3}=
$$
$$
=\nabla\int\mathbf{p'}f
\delta(\mathbf{p'}-\mathbf{p})d\,\mathbf{p'}=\mathbf{p}\nabla
f(\mathbf{r},\mathbf{p},t).
$$

Two following equalities can be verified similarly
$$
\iint\dfrac{e}{mc}\mathbf{A}(\mathbf{r},t)[\nabla f(\mathbf{r},
\mathbf{p}',t)]
e^{i(\mathbf{p'}-\mathbf{p})\mathbf{a}/\hbar}
\dfrac{d^3a\,d^3p'}{(2\pi\hbar)^3}=$$$$=
\dfrac{e}{mc}\mathbf{A}(\mathbf{r},t) \nabla f(\mathbf{r},\mathbf{p},t),
$$
and
$$
\iint\dfrac{e}{mc}[\div{\mathbf{A}(\mathbf{r},t)}] f(\mathbf{r},
\mathbf{p}',t)
e^{i(\mathbf{p'}-\mathbf{p})\mathbf{a}/\hbar}
\dfrac{d^3a\,d^3p'}{(2\pi\hbar)^3}=$$$$=
\dfrac{e}{mc}[\div{\mathbf{A}}(\mathbf{r},t)] f(\mathbf{r},\mathbf{p},t).
$$

Then the equation (1.6) can be rewritten as following
$$
\dfrac{\partial f}{\partial t}+\dfrac{1}{m}
\Big(\mathbf{p}-\dfrac{e}{c}\mathbf{A}\Big)\nabla f-\dfrac{e}{mc}
[\div{\mathbf{A}(\mathbf{r},t)}]f(\mathbf{r},\mathbf{p},t)=
W[f].
\eqno{(1.8)}
$$

In the equation (1.8) the symbol $W[f]$ is the Wigner
--- Vlasov integral, defined by the equality
$$
W[f]=
\iint\left\{
\dfrac{e}{2mc}
\Big[\mathbf{A}(\mathbf{r}+\dfrac{\mathbf{a}}{2},t)+
\mathbf{A}(\mathbf{r}-\dfrac{\mathbf{a}}{2},t)-
2\mathbf{A}(\mathbf{r},t)\Big]\nabla f\right.+
$$\smallskip
$$
+\dfrac{ie}{ mc\hbar}\Big[
\mathbf{A}(\mathbf{r}+\dfrac{\mathbf{a}}{2},t)
-\mathbf{A}(\mathbf{r}-\dfrac{\mathbf{a}}{2},t)\Big]\mathbf{p'}f+
$$\smallskip
$$
+\dfrac{e}{2mc}\Big[\div{\mathbf{A}(\mathbf{r}+
\dfrac{\mathbf{a}}{2},t)
}+\div{\mathbf{A}(\mathbf{r}-\dfrac{\mathbf{a}}{2},t)}-
2\div{\mathbf{A}(\mathbf{r},t)}\Big]f-
$$\smallskip
$$-
\dfrac{i e^2}{2 mc^2\hbar}\Big[\mathbf{A}^2(\mathbf{r}+
\dfrac{\mathbf{a}}{2},t)-
\mathbf{A}^2(\mathbf{r}-\dfrac{\mathbf{a}}{2},t)\Big]f-
$$\smallskip
$$
-\left. \dfrac{ie}{\hbar}\Big[U(\mathbf{r}+\dfrac{\mathbf{a}}{2},t)-
U(\mathbf{r}-\dfrac{\mathbf{a}}{2},t)\Big]f\right\}
e^{i(\mathbf{p'}-\mathbf{p})\mathbf{a}/\hbar}
\dfrac{d^3a\,d^3p'}{(2\pi\hbar)^3}.
\eqno{(1.9)}
$$
\smallskip

The energy of the particle is equal to
$$
\E=\E(\mathbf{r},\mathbf{p},t)=\dfrac{1}{2m}\Big(\mathbf{p}-
\dfrac{e}{c}\mathbf{A}\Big)^2+eU.
$$

Then the velocity of the particle $\mathbf{v}$ is equal to
$$
\mathbf{v}=\mathbf{v}(\mathbf{r},\mathbf{p},t)=
\dfrac{\partial \E}{\partial \mathbf{p}}=
\dfrac{1}{m}\Big(\mathbf{p}-\dfrac{e}{c}\mathbf{A}\Big),
$$
besides,
$$
\nabla \mathbf{v}=-\dfrac{e}{mc}\div{\mathbf{A}}.
$$

Hence, the left--hand part of the equation (1.9) equals to
$$
\dfrac{\partial f}{\partial t}+\dfrac{1}{m}
\Big(\mathbf{p}-\dfrac{e}{c}\mathbf{A}\Big)\nabla f-f\dfrac{e}{mc}
\div{\mathbf{A}}=
\dfrac{\partial f}{\partial t}+\mathbf{v}\nabla f+f\nabla \mathbf{A}=
$$
$$
=\dfrac{\partial f}{\partial t}+\nabla(\mathbf{v}f).
$$

Therefore the equation (1.9) can be rewritten in standard for
transport theory form
$$
\dfrac{\partial f}{\partial t}+\nabla (\mathbf{v}f)=W[f].
\eqno{(1.10)}
$$

In the case of collisional plasma we may write the kinetic equation
(1.10) as following
$$
\dfrac{\partial f}{\partial t}+\nabla (\mathbf{v}f)=B[f,f]+
W[f].
\eqno{(1.11)}
$$

In the equation (1.11) the symbol $B[f,f]$ represents the collision
integral.

Under the electron scattering on impurity we will consider the
equation (1.11) with collision integral in the form of relaxation
$\tau$--model \cite{BGK}, \cite{Opher}
$$
\dfrac{\partial f}{\partial t}+\nabla(\mathbf{v}f)=
\dfrac{f^{(0)}-f}{\tau}+W[f].
\eqno{(1.12)}
$$

In the equation (1.12) $\tau$ is the mean time between two
consecutive collisions,  $\tau=1/\nu$, $\nu$ is the effective collision
electron frequency with plasma particles,
$f^{(0)}$ is the local equilibrium Fermi --- Dirac
distribution function,
$$
f^{(0)}=f^{(0)}(\mathbf{r},\mathbf{p},t)=
\Big[1+\exp\Big(\dfrac{\E-\mu}{k_BT}\Big)\Big]^{-1}.
$$

Here  $k_B$ is the Boltzmann constant, $T$ is the plasma temperature,
$\E$ is the electron  energy,
$\mu$ is the chemical potential of electron gas.

In an explicit form the local equilibrium distribution function has
the following form
$$
f^{(0)}=f^{(0)}(\mathbf{r},\mathbf{p},t)=
\Bigg[1+\exp\Big[\dfrac{\big[\mathbf{p}-
(e/c)\mathbf{A}(\mathbf{r},t)\big]^2}{2mk_BT}+
\dfrac{eU(\mathbf{r},t)-\mu}{k_BT}\Big]\Bigg]^{-1}.
$$

We introduce the dimensionless electron velocity
$\mathbf{C}(\mathbf{r},\mathbf{p},t)$, scalar potential $\phi(\mathbf{r},t)$
and chemical potential $\alpha$
$$
\mathbf{C}(\mathbf{r},\mathbf{p},t)=
\dfrac{\mathbf{v}(\mathbf{r},\mathbf{p},t)}{v_T},\qquad
\phi(\mathbf{r},t)=\dfrac{eU(\mathbf{r},t)}{k_BT}, \qquad
\alpha=\dfrac{\mu}{k_BT},
$$
where
$v_T=\dfrac{1}{\sqrt{\beta}}$ is the thermal electron velocity,
$\beta=\dfrac{m}{2k_BT}$.

Now local equilibrium function can be presented in terms of the
electron velocity as follows
$$
f^{(0)}(\mathbf{r},\mathbf{p},t)=
\Big[1+\exp\Big(\dfrac{mv^2(\mathbf{r},\mathbf{p},t)}{2k_BT}+
\dfrac{eU(\mathbf{r},t)-\mu}{k_BT}\Big)
\Big]^{-1},
$$
or, in dimensionless parameters,
$$
f^{(0)}(\mathbf{r},\mathbf{p},t)=
\dfrac{1}{1+\exp\big[C^2(\mathbf{r},\mathbf{p},t)+
\phi(\mathbf{r},t)-\alpha\big]}.
\eqno{(1.13)}
$$

We designate \(\chi=\alpha-\phi\). Then we have
$$
f^{(0)}=\dfrac{1}{1+e^{C^2-\chi}}.
$$

The quantity $\chi $ is defined from the conservation law
of number of particles
$$
\int f d\Omega_F=
\int f^{(0)}d\Omega_F.
$$

Here $d\Omega_F$ is the quantum measure for electrons,
$$
d\Omega_F=\dfrac{2d^3p}{(2\pi\hbar)^3}.
$$

Let's note, that in the case of constant potentials $U={\rm const},
\mathbf{A}=\const$ the equilibrium distribution function (1.13) is
the solution of the equation (1.12).

Let's find the electron concentration (numerical density) $N$ and
mean electron velocity $\mathbf{u}$ in an equilibrium state. These
macroparameters are defined as follows
$$
N(\mathbf{r},t)=\int f(\mathbf{r},\mathbf{p},t)d\Omega_F,
$$
$$
\mathbf{u}(\mathbf{r},t)=\dfrac{1}{N(\mathbf{r},t)}
\int \mathbf{v}(\mathbf{r},\mathbf{p},t)
f(\mathbf{r},\mathbf{p},t)d\Omega_F.
$$

For calculation of these macroparameters in equilibrium condition it
is necessary to put $f=f^{(0)}$, where $f^{(0)}$ is defined by
equality (1.13). We designate these macroparameters in equilibrium
condition through $N^{(0)}(\mathbf{r},t)$ and
$\mathbf{u}^{(0)}(\mathbf{r},t)$.

Let's carry out the replacement of the integration variable
$$
\mathbf{p}-\dfrac{e}{c}\mathbf{A}(\mathbf{r},t)=\mathbf{p}'
$$
in the previous equalities. Then, passing to integration in spherical
co\-or\-di\-na\-tes, for numerical density in an equilibrium state
we get
$$
N^{(0)}=\dfrac{m^3v_T^3}{\pi^2\hbar^3}f_2(\alpha-\phi),
\eqno{(1.14)}
$$
where
$$
f_2(\alpha-\phi)=\int\limits_{0}^{\infty}\dfrac{x^2\;dx}{1+\exp(x^2+
\phi-\alpha)}=\int\limits_{0}^{\infty}x^2f_F(\alpha-\phi)\,dx.
$$

In the same way, as for numerical density, for mean velocity in
equilibrium state we derive
$$
\mathbf{u}^{(0)}(\mathbf{r},t)=\dfrac{1}{N^{(0)}}
\int \mathbf{v}(\mathbf{r},\mathbf{p},t)
f^{(0)}(\mathbf{r},\mathbf{p},t)d\Omega_F,
$$
or, in explicit form,
$$
\mathbf{u}^{(0)}(\mathbf{r},t)=\dfrac{2}{N^{(0)}(2\pi\hbar)^3}\int
\dfrac{[\mathbf{p}-(e/c)\mathbf{A}]\;d^3p}
{1+\exp\Big[\dfrac{(\mathbf{p}-(e/c)\mathbf{A})^2}{2k_BTm}+
\dfrac{eU-\mu}{k_Tm}\Big]}.
$$

After the same change of variables
$\mathbf{p}-(e/c)\mathbf{A}(\mathbf{r},t)= \mathbf{p}'$ we receive
$$
\mathbf{u}^{(0)}(\mathbf{r},t)=
\dfrac{2}{N^{(0)}(2\pi\hbar)^3}\int \dfrac{\mathbf{p}'\;d^3p'}
{1+\exp\Big[\dfrac{{p'}^2}{2k_BTm}+\dfrac{eU-\mu}{k_Tm}\Big]}=0.
\eqno{(1.15)}
$$

So, the electron velocity in an equilibrium state according to
$(1.15)$ is equal to zero.

Let's note, that numerical electron density and their mean velocity
satisfy the usual continuity equation
$$
\dfrac{\partial N}{\partial t}+\mathbf{div}(N\mathbf{u})=0.
\eqno{(1.16)}
$$

For the derivation of the continuity equation (1.16) is necessary
to integrate the kinetic equation (1.12) by quantum measure for
electrons $d\Omega_F$ and to use the definition of numerical density
and mean velocity.
Then is necessary to use the conservation law  of number of
particles and to check up, if the integral by quantum measure
$d\Omega_F$ of Wigner --- Vlasov integral is equal to zero. Indeed,
we have
$$
\int W[f]\dfrac{2\;d^3p}{(2\pi\hbar)^3}=2\iint
\Big\{\cdots\Big\}
e^{i\mathbf{p'}\mathbf{a}/\hbar}\delta(\mathbf{a})\,d^3a\,d^3p'=
$$
$$
=2\int \Big\{\cdots\Big\}\Bigg|_{\mathbf{a}=0}d^3p\equiv 0,
$$
as after some algebra,
$$
 \Big\{\cdots\Big\}\Bigg|_{\mathbf{a}=0}\equiv 0.
$$

Here the symbol $\{\cdots\}$ means the same expression, as in the
right-hand side of the equation (1.9).

Let's note, that the left-hand side of the kinetic equation (1.11)
or (1.12) takes standard form for transport theory under the
following gauge condition:
$$
\div{\mathbf{A}(\mathbf{r},t)}=0.
\eqno{(1.17)}
$$

Thus, i.e. in case of gauge (1.17), the kinetic equation has
the following form
$$
\dfrac{\partial f}{\partial t}+\mathbf{v}\nabla f=B[f,f]+
W[f].
\eqno{(1.18)}
$$

Here the Wigner --- Vlasov integral equals to
$$
W[f]=
\iint\left\{
\dfrac{e}{2mc}
\Big[\mathbf{A}(\mathbf{r}+\dfrac{\mathbf{a}}{2},t)+
\mathbf{A}(\mathbf{r}-\dfrac{\mathbf{a}}{2},t)-
2\mathbf{A}(\mathbf{r},t)\Big]\nabla f\right.+
$$
\hspace{0.5cm}
$$
+\dfrac{ie}{ mc\hbar}\Big[
\mathbf{A}(\mathbf{r}+\dfrac{\mathbf{a}}{2},t)
-\mathbf{A}(\mathbf{r}-\dfrac{\mathbf{a}}{2},t)\Big]\mathbf{p'}
f-$$$$-\dfrac{i e^2}{2 mc^2\hbar}
\Big[\mathbf{A}^2(\mathbf{r}+\dfrac{\mathbf{a}}{2},t)-
\mathbf{A}^2(\mathbf{r}-\dfrac{\mathbf{a}}{2},t)\Big]f-
$$
\hspace{0.5cm}
$$
-\left. \dfrac{ie}{\hbar}\Big[U(\mathbf{r}+\dfrac{\mathbf{a}}{2},t)-
U(\mathbf{r}-\dfrac{\mathbf{a}}{2},t)\Big]f\right\}
e^{i(\mathbf{p'}-\mathbf{p})\mathbf{a}/\hbar}
\dfrac{d^3a\,d^3p'}{(2\pi\hbar)^3}.
\eqno{(1.19)}
$$\\

\begin{center}
  {\bf 3. LINEARIZATION OF THE KINETIC EQUATION AND ITS SOLUTION}
\end{center}

Let's consider the kinetic equation with collision integral in the
form  of $\tau$--model and suppose, that the scalar potential is
equal to zero $U(\mathbf {r},t)\equiv 0$.

We take vector potential which is orthogonal to the direction of the
wave vector $\mathbf {k}:\mathbf{k}\mathbf{A}=0$ in the form of
a running harmonic wave
$$
\mathbf{A}(\mathbf{r},t)=\mathbf{A}_0
e^{i(\mathbf{k}\mathbf{r}-\omega t)}.
$$

We suppose that the vector potential is small enough. This
assump\-tion allows us to linearize the equation and to neglect terms
quadratic in electric field.

Then the equation (1.18) can be reduced to
$$
\dfrac{\partial f}{\partial t}+\mathbf{v}\nabla f=
\dfrac{f^{(0)}-f}{\tau}+W[f].
\eqno{(2.1)}
$$

In this case chemical potential is equal to a constant.

In the equation (2.1) local equilibrium Fermi --- Dirac distribution
is simplified as following
$$
f^{(0)}=f^{(0)}(\mathbf{r},\mathbf{p},t)=
\Big[1+\exp\Big(C^2(\mathbf{r},\mathbf{p},t)-\alpha\Big)\Big]^{-1}.
\eqno{(2.2)}
$$

The Wigner --- Vlasov integral (1.19) also can be simplified
essentially and has the following form
$$
W[f]=\dfrac{ie}{mc\hbar }
\iint
\Big[\mathbf{A}(\mathbf{r}+\dfrac{\mathbf{a}}{2},t)
-\mathbf{A}(\mathbf{r}-\dfrac{\mathbf{a}}{2},t)\Big]\mathbf{p'}
f^{}
e^{i(\mathbf{p'}-\mathbf{p})\mathbf{a}/\hbar}
\dfrac{d^3a\,d^3p'}{(2\pi\hbar)^3}.
\eqno{(2.3)}
$$

We notice, that
$$
\mathbf{A}(\mathbf{r}+\dfrac{\mathbf{a}}{2},t)
-\mathbf{A}(\mathbf{r}-\dfrac{\mathbf{a}}{2},t)=\mathbf{A}(\mathbf{r},t)
\Big[e^{i\mathbf{k}\mathbf{a}/2}-e^{-i\mathbf{k}\mathbf{a}/2}\Big].
$$

Calculating the integral in (2.3), we find that
$$
W[f]=\dfrac{ie}{mc\hbar}\mathbf{A}(\mathbf{r},t)
\iint \Big[e^{i\mathbf{k}\mathbf{a}/2}-e^{-i\mathbf{k}\mathbf{a}/2}\Big]
e^{i(\mathbf{p}'-\mathbf{p})\mathbf{a}/\hbar}\dfrac{d^3a\,d^3p'}
{(2\pi\hbar)^3}.
$$

The internal integral is equal to
$$
\dfrac{1}{(2\pi\hbar)^3}\int \Big\{\exp\Big(i\Big[\mathbf{p}'-\mathbf{p}+
\frac{\mathbf{k}\hbar}{2}\Big]\dfrac{\mathbf{a}}{\hbar}\Big)-
\exp\Big(i\Big[\mathbf{p}'-\mathbf{p}+
\frac{\mathbf{k}\hbar}{2}\Big]\dfrac{\mathbf{a}}{\hbar}\Big)\Big\}d^3a=
$$
$$
=\delta\Big(\mathbf{p}'-\mathbf{p}+\dfrac{\hbar\mathbf{k}}{2}\Big)-
\delta\Big(\mathbf{p}'-\mathbf{p}-\dfrac{\hbar\mathbf{k}}{2}\Big).
$$

We calculate the Wigner --- Vlasov integral
$$
W[f]=
$$
$$=\mathbf{A}(\mathbf{r},t)\dfrac{ie}{mc\hbar}
\int \Big[\delta(\mathbf{p}'-\mathbf{p}+\dfrac{\hbar\mathbf{k}}{2})-
\delta(\mathbf{p}'-\mathbf{p}-\dfrac{\hbar\mathbf{k}}{2})\Big]
\mathbf{p}'f(\mathbf{r},\mathbf{p}',t)\,d^3p'=
$$\vspace{0.3cm}
$$
=\mathbf{A}(\mathbf{r},t)\dfrac{ie}{mc\hbar}
\Big[\Big(\mathbf{p}-\dfrac{\hbar \mathbf{k}}{2}\Big)f(\mathbf{r},
\mathbf{p}-\dfrac{\hbar \mathbf{k}}{2},t)-
\Big(\mathbf{p}+\dfrac{\hbar \mathbf{k}}{2}\Big)f(\mathbf{r},
\mathbf{p}+\dfrac{\hbar \mathbf{k}}{2},t)\Big]=
$$\vspace{0.3cm}
$$
=\mathbf{A}(\mathbf{r},t)\dfrac{ie}{mc\hbar}
\Big\{\mathbf{p}\Big[f(\mathbf{r},
\mathbf{p}-\dfrac{\hbar \mathbf{k}}{2},t)-f(\mathbf{r},
\mathbf{p}+\dfrac{\hbar \mathbf{k}}{2},t)\Big]-
$$
\vspace{0.3cm}
$$
-\dfrac{\hbar\mathbf{k}}{2}\Big[f(\mathbf{r},
\mathbf{p}-\dfrac{\hbar \mathbf{k}}{2},t)+f(\mathbf{r},
\mathbf{p}+\dfrac{\hbar \mathbf{k}}{2},t)\Big]\Big\}=
$$\vspace{0.3cm}
$$
=\mathbf{A}(\mathbf{r},t)\dfrac{ie}{mc\hbar}\mathbf{p}\Big(f_+-f_-\Big),
$$
where
$$
f_{\pm}\equiv f(\mathbf{r},\mathbf{p}\mp\dfrac{\hbar
\mathbf{k}}{2},t).
$$

Consequently, the Wigner --- Vlasov integral is equal to
$$
W[f]=\dfrac{ie}{mc\hbar}\mathbf{p}\mathbf{A}
(\mathbf{r},t)\Big[f_+^{}-f_-^{}\Big]=$$$$\hspace{2cm}+
\dfrac{iep_T}{mc\hbar}\mathbf{P}\mathbf{A}(\mathbf{r},t)
\Big[f_+^{}-f_-^{}\Big]=$$$$\hspace{4cm}=
\dfrac{iev_T}{c\hbar}\mathbf{P}\mathbf{A}
(\mathbf{r},t)\Big[f_+^{}-f_-^{}\Big].
\eqno{(2.4)}
$$

Here and below the expression $\mathbf{P}\mathbf{A}$ means
scalar production.

Further we will use dimensionless velocity $\mathbf{C}$ in the form
$$
\mathbf{C}=\dfrac{\mathbf{v}}{v_T}=\dfrac{\mathbf{p}}{p_T}-
\dfrac{e}{cp_T}\mathbf{A}(\mathbf{r},t)\equiv \mathbf{P}-
\dfrac{e}{cp_T}\mathbf{A}(\mathbf{r},t),
$$
where $\mathbf{P}=\dfrac{\mathbf{p}}{p_T}$ is the dimensionless momentum.

Then
$$
\mathbf{v}=\mathbf{v}(\mathbf{r}, \mathbf{P},t)=
v_T\Big(\mathbf{P}-\dfrac{e}{cp_T}\mathbf{A}(\mathbf{r},t)\Big).
$$

In linear approximation is possible to replace the function $f$
in Wigner --- Vlasov integral by the absolute Fermi --- Dirac
distribution, i.e. we put $f=f_F(P)$, where
$$
f_F(P)=\dfrac{1}{1+\exp (P^2-\alpha)}, \qquad \alpha=\const.
$$

Here Wigner --- Vlasov integral (2.4) has the following form
$$
W[f_F]=\dfrac{iev_T}{c\hbar}\mathbf{P}\mathbf{A}
(\mathbf{r},t)\Big[f_F^+-f_F^-\Big],
$$
where
$$
f_F^{\pm}\equiv f_F^{\pm}(\mathbf{P})=
\dfrac{1}{1+
\exp\Big[\Big(\mathbf{P}\mp \dfrac{\hbar \mathbf{k}}{2p_T}\Big)^2
-\alpha\Big]},
$$
and $p_T=mv_T$ is the thermal electron momentum, or,
$$
f_F^{\pm}=\dfrac{1}{1+e^{P^2_{\pm}-\alpha}}.
$$

Here
$$
P^2_{\pm}=\Big(\mathbf{P}\mp\dfrac{\hbar \mathbf{k}}{2p_T}\Big)^2
=\Big(P_x\mp\dfrac{\hbar k_x}{2p_T}\Big)^2+
\Big(P_y\mp\dfrac{\hbar k_y}{2p_T}\Big)^2+
\Big(P_z\mp\dfrac{\hbar k_z}{2p_T}\Big)^2,
$$
or
$$
P^2_{\pm}=\dfrac{\Big(p_x\mp\dfrac{\hbar k_x}{2}\Big)^2+
\Big(p_y\mp\dfrac{\hbar k_y}{2}\Big)^2+
\Big(p_z\mp\dfrac{\hbar k_z}{2}\Big)^2}{p_T^2}.
$$

The linearization of the Wigner equilibrium  function (2.2) we will
carry out in terms of vector potential $\mathbf{A}(\mathbf{r},t)$
$$
f^{(0)}=f^{(0)}\Big|_{\mathbf{A}=0}+
\dfrac{\partial f^{(0)}}
{\partial \mathbf{A}}\Bigg|_{\mathbf{A}=0}
 \mathbf{A}(\mathbf{r},t),
$$
or, in explicit form
$$
f^{(0)}=f_F(P)+g(P)\dfrac{2e}{cp_T}\mathbf{P}\mathbf{A}(\mathbf{r},t),
\eqno{(2.5)}
$$
$$
g(P)=\dfrac{e^{P^2-\alpha}}{(1+e^{P^2-\alpha})^2}.
$$

Considering decomposition (2.5), we will search for Wigner's function
in the form
$$
f=f_F(P)+g(P)\dfrac{2e}{cp_T}\mathbf{P}\mathbf{A}(\mathbf{r},t)+
g(P)(\mathbf{P} \mathbf{A}(\mathbf{r},t))h(\mathbf{P}).
\eqno{(2.6)}
$$

We receive the following equation
$$
\big[\mathbf{P}\mathbf{A}(\mathbf{r},t)\big] \;g(P)(\nu-i\omega+
i\mathbf{k} \mathbf{v})h(\mathbf{P})=
$$
$$
=\dfrac{iev_T}{c\hbar}
\big[\mathbf{P}\mathbf{A}(\mathbf{r},t)\big](f_F^+-f_F^-)+
\dfrac{2ie}{cp_T}g(P)(\omega-v_T\mathbf{k}\mathbf{P})
\big[\mathbf{P}\mathbf{A}(\mathbf{r},t)\big].
$$

From this equation we find
$$
\big[\mathbf{P}\mathbf{A}(\mathbf{r},t)\big]g(P)h(\mathbf{P})=
\big[\mathbf{P}\mathbf{A}(\mathbf{r},t)\big]\dfrac{2ie}{cp_T}
\Bigg[\dfrac{\omega-v_T \mathbf{k} \mathbf{P}}
{\nu-i\omega+iv_T\mathbf{k} \mathbf{P}}g(P)+
$$
$$
+
\dfrac{mv_T^2}{2\hbar}\dfrac{f_F^+(\mathbf{P})-f_F^-(\mathbf{P})}
{\nu-i\omega+iv_T\mathbf{k} \mathbf{P}}\Bigg].
\eqno{(2.7)}
$$

With the help of (2.6) and (2.7) we construct the full distribution
function
$$
f=f^{(0)}+g(P)h(\mathbf{P})\mathbf{P A}=
$$
$$
=f^{(0)}+\dfrac{2ie}{cp_T}\mathbf{P A}\Bigg[\dfrac{\omega-v_T\mathbf{k P}}
{\nu-i \omega +iv_T\mathbf{k P}}g(P)+
\dfrac{mv_T^2}{2\hbar}\dfrac{f_F^+(\mathbf{P})-f_F^-(\mathbf{P})}
{\nu-i \omega +iv_T\mathbf{k P}}\Bigg],
$$  \medskip
or
$$
f=f^{(0)}+\dfrac{2ie}{cp_T}\mathbf{P A}\Bigg[
\dfrac{\omega\tau-\mathbf{k}_1\mathbf{P}}
{1-i \omega\tau +i\mathbf{k}_1\mathbf{P}}g(P)+
\dfrac{mv_T^2}{2\hbar \nu}\dfrac{f_F^+(\mathbf{P})-f_F^-(\mathbf{P})}
{1-i \omega\tau +i\mathbf{k}_1\mathbf{P}}\Bigg].
\eqno{(2.8)}
$$ \medskip

Here $\mathbf{k}_1=\mathbf{k}l$, $l$ is the electron mean free path,
$\;l=v_T\tau$, $\mathbf{k}_1$ is the dimensionless wave vector.

We consider the connection between electric field and potentials
$$
\mathbf{E}(\mathbf{r},t)=
-\dfrac{1}{c}\dfrac{\partial \mathbf{A}(\mathbf{r},t)}{\partial t}-
\dfrac{\partial U(\mathbf{r},t)}{\partial \mathbf{r}},
$$
or
$$
\mathbf{E}(\mathbf{r},t)=\dfrac{i \omega}{c}\mathbf{A}(\mathbf{r},t).
$$

Hence,  the current is connected with vector potential as
$$
\mathbf{j}(\mathbf{r},t)=
\sigma_{tr}\dfrac{i \omega}{c}\mathbf{A}(\mathbf{r},t).
$$

By definition, the current is equal to
$$
\mathbf{j}(\mathbf{r},t)=
e\int \mathbf{v}(\mathbf{r},\mathbf{p},t)f\dfrac{2\,d^3p}{(2\pi\hbar)^3}.
$$

Let's note, that the current in the equilibrium state  is equal to zero
$$
\mathbf{j}^{(0)}(\mathbf{r},t)=
e\int \mathbf{v}(\mathbf{r},\mathbf{P},t)
f^{(0)}\dfrac{2p_T^3\,d^3P}{(2\pi\hbar)^3}=0.
$$

Indeed, considering that mean electron velocity in the equilibrium
state is equal to zero, according to $(1.15)$ we have
$$
\mathbf{j}^{(0)}(\mathbf{r},t)=
e N^{(0)}\mathbf{u}^{(0)}(\mathbf{r},t)\equiv 0.
$$

Hence, with the use of equality (2.8) we have the following equality
$$
\mathbf{j}(\mathbf{r},t)=i\dfrac{4e^2p_T^2}{(2\pi\hbar)^3c}
\int \Big(\mathbf{P A}\Big)\mathbf{v}(\mathbf{r},\mathbf{P},t)\times
$$
$$
\times\Bigg[\dfrac{\omega\tau-\mathbf{k}_1\mathbf{P}}
{1-i \omega\tau +i\mathbf{k}_1 \mathbf{P}}\,g(P)+
\dfrac{\E_T}{\hbar \nu}\dfrac{f_F^+(\mathbf{P})-f_F^-(\mathbf{P})}
{1-i \omega\tau +i\mathbf{k}_1 \mathbf{P}}\Bigg]d^3P,
$$
where $\E_T$ is the thermel kinetic energy of  electrons,
$$
\E_T=\dfrac{mv_T^2}{2}.
$$

Substituting obvious expression for the velocity into this equality
$$
\mathbf{v}(\mathbf{r},\mathbf{P},t)=
\dfrac{\mathbf{p}}{m}-\dfrac{e \mathbf{A}(\mathbf{r},t)}{mc}=
\dfrac{p_T\mathbf{P}}{m}-\dfrac{e \mathbf{A}(\mathbf{r},t)}{mc},
$$
and,  after linearization of it by vector field, we receive
$$
\mathbf{j}(\mathbf{r},t)=
i\dfrac{4e^2p_T^3}{(2\pi \hbar)^3mc}\int \Big(\mathbf{P\,A}(\mathbf{r},t)
\Big)\mathbf{P}\times
$$
$$
\times\Bigg[
\dfrac{\omega\tau-\mathbf{k}_1 \mathbf{P}}
{1-i \omega\tau +i\mathbf{k}_1 \mathbf{P}}g(P)+
\dfrac{\E_T}{\hbar \nu}\dfrac{f_F^+(\mathbf{P})-f_F^-(\mathbf{P})}
{1-i \omega\tau +i\mathbf{k}_1 \mathbf{P}}\Bigg]d^3P.
$$

It is seen easily, that all the components of the vector
$\mathbf{j}$, which are orthogonal to the vector $\mathbf{A}$ are
equal to zero. Therefore
$$
\mathbf{j}(\mathbf{r},t)=i
\dfrac{4e^2p_T^3\mathbf{A}(\mathbf{r},t)}{(2\pi \hbar)^3mc}
\int \Big(\mathbf{e}_1
\mathbf{P}\Big)^2\Bigg[
\dfrac{\omega\tau-\mathbf{k}_1 \mathbf{P}}
{1-i \omega\tau +i\mathbf{k}_1 \mathbf{P}}g(P)+$$$$+
\dfrac{\E_T}{\hbar \nu}\dfrac{f_F^+(\mathbf{P})-f_F^-(\mathbf{P})}
{1-i \omega\tau +i\mathbf{k}_1 \mathbf{P}}\Bigg]d^3P.
$$

Here $\mathbf{e}_1=\mathbf{A}/A $ is the unit vector directed
lengthwise $\mathbf{A}$. In view of the symmetry the value of
integral will not change, if the vector $\mathbf{e}_1$ is replaced
by any other unit vector $\mathbf{e}_2$, perpendicular to the vector
$\mathbf{k}_1$. Therefore
$$
\int \Big(\mathbf{e}_1
\mathbf{P}\Big)^2[s]d^3P=
\int \Big(\mathbf{e}_2
\mathbf{P}\Big)^2[s]d^3P=
$$
$$
=\dfrac{1}{2}\int \Bigg[\Big(\mathbf{e}_1 \mathbf{P}\Big)^2
+\Big(\mathbf{e}_2 \mathbf{P}\Big)^2\Bigg]
[s]d^3P,
$$
where
$$\qquad \qquad \mathbf{e}_2=\dfrac{\mathbf{A} \times \mathbf{k}_1}
{|\mathbf{A} \times \mathbf{k}_1|}=
\dfrac{\mathbf{A} \times \mathbf{k}_1}{Ak_1},
$$
and $\mathbf{A}\times\mathbf{k}_1$ is the vector product.

But we have further
$$
\Big(\mathbf{e}_1 \mathbf{P}\Big)^2
+\Big(\mathbf{e}_2 \mathbf{P}\Big)^2=P^2-
\dfrac{(\mathbf{P\mathbf{k}_1)^2}}{k_1^2}=
$$$$=P^2-(\mathbf{Pn})^2 \equiv P_\perp^2,
$$
where $\mathbf{n}$ is the unit vector directed along
the vector $\mathbf{k}_1,\; \mathbf{n}=\dfrac{\mathbf{k}_1}{k_1}$.

Hence for the current density we receive the following expression
$$
\mathbf{j}(\mathbf{r},t)=i
\dfrac{2e^2p_T^3\mathbf{A}(\mathbf{r},t)}{(2\pi \hbar)^3mc}\int
\Bigg[
\dfrac{\omega\tau-\mathbf{k}_1 \mathbf{P}}
{1-i \omega\tau +i\mathbf{k}_1 \mathbf{P}}g(P)+
$$
$$
+\dfrac{\E_T}{\hbar \nu}\dfrac{f_F^+(\mathbf{P})-f_F^-(\mathbf{P})}
{1-i \omega\tau +i\mathbf{k}_1 \mathbf{P}}\Bigg]P_\perp^2d^3P.
$$

Replacing the current in the left--hand side of this equality by the
expression in terms of field, we receive:
$$
\sigma_{tr}\dfrac{i \omega}{c}\mathbf{A}(\mathbf{r},t)=
i\dfrac{2e^2p_T^3\mathbf{A}(\mathbf{r},t)}{(2\pi \hbar)^3mc}\int \times
$$
$$
\times\Bigg[
\dfrac{\omega\tau-\mathbf{k}_1 \mathbf{P}}
{1-i \omega\tau +i\mathbf{k}_1 \mathbf{P}}g(P)+
\dfrac{\E_T}{\hbar\nu}\dfrac{f_F^+(\mathbf{P})-f_F^-(\mathbf{P})}
{1-i \omega\tau +i\mathbf{k}_1 \mathbf{P}}\Bigg]P_\perp^2 d^3P.
$$\medskip

\begin{center}
  \bf 4.  ELECTRIC CONDUCTIVITY AND DIELECTRIC FUNCTION
\end{center}

From the last formula we receive the following expression for the
transverse dielectric function in quantum plasma
$$
\sigma_{tr}=
\dfrac{2e^2p_T^3}{(2\pi \hbar)^3m\omega}\int \Big[
P^2-
\dfrac{(\mathbf{P}\mathbf{k}_1)^2}{k_1^2}\Big]\Bigg[
\dfrac{\omega\tau-\mathbf{k}_1 \mathbf{P}}
{1-i \omega\tau +i\mathbf{k}_1 \mathbf{P}}g(P)+
$$
$$
+\dfrac{\E_T}{\hbar\nu}\dfrac{f_F^+(\mathbf{P})-f_F^-(\mathbf{P})}
{1-i \omega\tau +i\mathbf{k}_1 \mathbf{P}}\Bigg]d^3P.
$$

We will transform expression for transverse  conductivity and
we will bring it to the form
$$
\sigma_{tr}=
\dfrac{2e^2p_T^3}{(2\pi \hbar)^3 m \omega}\int \Bigg[
(\omega\tau-\mathbf{k}_1\mathbf{P})g(P)
+\dfrac{\E_T}{\hbar \nu}(f_F^+-f_F^-)\Bigg]
\dfrac{P_\perp^2d^3P}{1-i \omega\tau +i\mathbf{k}_1
\mathbf{P}}.
$$

With the use of the equality (1.14)  we will present the previous
formula in the form
$$
\sigma_{tr}=\dfrac{\sigma_0}{4\pi f_2(\alpha)}
\int\Bigg[\big[1-(\omega\tau)^{-1}\mathbf{k}_1\mathbf{P}\Big]g(P)+
\hspace{4cm}
$$
$$\hspace{2cm}
+\dfrac{\E_T}{\hbar \omega}
\Big[f_F^+(\mathbf{P})-f_F^-(\mathbf{P})\Big]\Bigg]
\dfrac{P_\perp^2\;d^3P}{1-i\omega\tau+i\mathbf{k}_1
\mathbf{P}}.
\eqno{(3.1)}
$$

Here the function $f_2(\alpha)$ has been entered above and in the
absence of the scalar potential it is defined by equality
$$
f_2(\alpha)=\int\limits_{0}^{\infty}x^2f_F(x)dx=\int\limits_{0}^{\infty}
\dfrac{x^2\,dx}{1+e^{x^2-\alpha}}=\dfrac{1}{2}\int\limits_{0}^{\infty}
\ln(1+e^{\alpha-x^2})dx.
$$

The quantity $\sigma_0$ is defined by classical expression for the
static electric conductivity
$$
\sigma_0=\dfrac{e^2N^{(0)}}{m\nu}.
$$

Dielectric function we will find according to the formula
$$
\varepsilon_{tr}=
1+\frac{4\pi i}{\omega}\sigma_{tr}.
$$

Substituting electric conductivity (3.1) into this equality, we
receive the expression for dielectric permittivity in quantum
collision plasma
$$
\varepsilon_{tr}=1+\dfrac{\omega_p^2}{\omega^2}
 \dfrac{i}{4\pi f_2(\alpha)}
\int\Bigg\{\Big[\omega\tau-\mathbf{k}_1\mathbf{P}\Big]g(P)+
\hspace{4cm}
$$
$$\hspace{2cm}
+\dfrac{\E_T}{\hbar
\nu}\Big[f_F^+(\mathbf{P})-f_F^-(\mathbf{P})\Big]\Bigg\}
\dfrac{P_\perp^2\;d^3P}{1-i\omega \tau+i\mathbf{k}_1\mathbf{P}}.
$$

We investigate some special cases of electroconductivity. In the
long-wave limit (when $k\to 0$) from (3.1) we receive the well known
classical expression
$$
\sigma_{tr}(k=0)=\sigma_0\dfrac{\nu}{\nu-i \omega}=
\dfrac{\sigma_0}{1-i\omega\tau}.
$$

Let's consider the quantum mechanical limit of the conductivity in
the case of arbitrary values of wave number, i.e. conductivity limit
in the case, when Planck's constant $\hbar\to 0$, and the quantity
$k$ is arbitrary.

Now we consider the case, when values of the wave number are
arbitrary, but Planck's constant converges to zero: $\hbar\to 0$.

When the values of $\hbar$ are small we have
$$
f_0^{\pm}(\mathbf{P})=f_F(P)\pm g(P)2\mathbf{P}\dfrac{\hbar
\mathbf{k}}{2mv_T},
$$
hence
$$
f_F^+(\mathbf{P})-f_F^-(\mathbf{P})=2g(P)2\mathbf{P}\dfrac{\hbar
\mathbf{k}}{2mv_T}.
$$

Therefore
$$
(\omega-v_T\mathbf{k}\mathbf{P})g(P)
+\dfrac{p_T^2}{2m\hbar}\big[f_F^+(\mathbf{P})-f_F^-(\mathbf{P})\big]=
\omega g(P).
$$

Thus, in linear approximation at small $\hbar$ (independently of the
quantity $k$) for transverse conductivity we receive
$$
\sigma_{tr}=\sigma_{tr}^{\rm classic},
$$
where
$$
\sigma_{tr}^{\rm classic}=\dfrac{\sigma_0}{4\pi f_2(\alpha)}\int
\dfrac{g(P)P_\perp^2\,d^3P}{1-i\omega \tau +i
\mathbf{k}_1\mathbf{P}}.
\eqno{(3.2)}
$$

The expression (3.2) accurately coincides with the expression of the
transverse conductivity for classical plasma with arbitrary
temperature.

Let's return to the expression (3.1). We present it in the form of
the sum of two components
$$
\sigma_{tr}=\sigma_{tr}^{\rm classic}+\sigma_{tr}^{\rm quant},
\eqno{(3.3)}
$$
where $\sigma^{\rm classic}$ is defined by the equality (3.2), and
second component $\sigma_{tr}^{\rm quant}$ corresponds to quantum
properties of the plasma under consideration
$$
\sigma_{tr}^{\rm quant}=\dfrac{\sigma_0}{4\pi f_2(\alpha)}
\int\Bigg[-\dfrac{\mathbf{k}_1\mathbf{P}}{\omega\tau}g(P)
+\hspace{5cm}$$$$\hspace{3cm}+
\dfrac{\E_T}{\hbar \omega}[f_F^+(\mathbf{P})-f_F^-(\mathbf{P})]\Bigg]
\dfrac{P_\perp^2\,d^3P}{1-i\omega\tau+i\mathbf{k}_1\mathbf{P}}.
\eqno{(3.4)}
$$

The quantum summand $\sigma_{tr}^{\rm quant}$ we will present in the
form, proportional to a square of the Planck's constant $\hbar$.

For this aim we use cubic expansion of $\sigma_{tr}^{\rm quant}$
by powers of $\hbar$. We will remind, that in linear approximation
by $\hbar$, as it was already specified, the quantity
$\sigma_{tr}^{\rm quant}$ disappears.
We will direct an axis $x$ along the wave vector $\mathbf{k}$.

Let's expand the Fermi --- Dirac distribution by degrees of
dimensionless wave number $q=\dfrac{k}{k_T}=\dfrac{k_1\hbar\nu}
{mv_T^2}$, where $k_T=\dfrac{p_T}{\hbar}$ is the thermal wave
number. We receive
$$
f_F^{\pm}(\mathbf{P})=f_F(P)\pm g(P)P_x q-
\Big[g'_{P^2}(P)P_x^2+\dfrac{1}{2}
g(P)\Big]\dfrac{q^2}{2}\pm
$$
$$
\pm\Big[g''_{P^2P^2}(P)P_x^2+
\dfrac{3}{2}g'_{P^2}(P)\Big]P_x\dfrac{q^3}{6}+\cdots.
$$

Here
$$
g'_{P^2}(P)=g'(P^2),\qquad\;g''_{P^2P^2}(P)=g''(P^2),
$$
$$
g'(P^2)=g(P)\dfrac{1-e^{P^2-\alpha}}{1+e^{P^2-\alpha}},
$$
$$
g''(P^2)=g(P)\Big[\Big(\dfrac{1-e^{P^2-\alpha}}{1+e^{P^2-\alpha}}\Big)^2-
2g(P)\Big].
$$

Now we will find the difference
$$
f_F^+(\mathbf{P})-f_F^-(\mathbf{P})=
2g(P)P_xq+\Big[g''(P^2)P_x^2+
\dfrac{3}{2}g'(P^2)\Big]P_x\dfrac{q^3}{3}+\cdots.
$$

By means of this expression we find, that
$$
-\dfrac{k_1P_x}{\omega\tau}g(P)+\dfrac{\E_T}{\hbar\omega}
[f_F^+(\mathbf{P})-f_F^-(\mathbf{P})]=
G(\mathbf{P})\dfrac{k_1^3\hbar^2 \nu^3}{6\omega
m^2v_T^4}+\cdots,
$$
where
$$
G(\mathbf{P})=P_x\Big[g''(P^2)P_x^2+
\dfrac{3}{2}g'(P^2)\Big].
$$

Substituting this expression into (3.4), we obtain, that the quantum
summand is proportional to the square of Planck's constant and it is
defined by expression
$$
\sigma_{tr}^{\rm quant}=\hbar^2 \sigma_0
\dfrac{k_1^3 \nu^3}{24\pi \omega m^2v_T^4 f_2(\alpha)}
\int\dfrac{G(\mathbf{P})(P^2-P_x^2)\,d^3P}{1-i\omega\tau+ik_1P_x}.
\eqno{(3.5)}
$$

In the expressions for classical and quantum components of the
conductivity we can simplify several integrals.

We break the triple integral to external one--dimensional integration
by the variable $P_x$ from $-\infty$ to $+\infty$ and internal
double integration by plane orthogonal to the axis $P_x$ in the
expression (3.3). The internal integration we carry out in polar
coordinates. Here we obtain that
$$
P^2=P_x+P_{\perp}^2, \qquad d^3P=dP_x\;d{\mathbf{P_{\perp}}}, \qquad
d{\mathbf{P_{\perp}}}=P_{\perp}\,dP_{\perp}d\chi,
$$
where $P_{\perp}$ is the polar radius, and $\chi$ is the polar angle.

Thus we receive, that
$$
\sigma_{tr}^{\rm classic}=\dfrac{\sigma_0}{4\pi f_2(\alpha)}
\int\limits_{-\infty}^{\infty}dP_x
\int\limits_{0}^{\infty}\int\limits_{0}^{2\pi}
\dfrac{g(P)\,P_{\perp}^3 dP_{\perp} d\chi}{1-i\omega \tau
+ik_1P_x},
$$
where
$$
g(P)=\dfrac{e^{P_x^2+P_{\perp}-\alpha}}{(1+e^{P_x^2+P_{\perp}-\alpha})^2}.
$$

Internal double integral we calculate in polar coordinates
$$
\int\limits_{0}^{\infty}\int\limits_{0}^{2\pi}
g(P)\,P_{\perp}^3 dP_{\perp} d\chi=2\pi \int\limits_{0}^{\infty}
\dfrac{P_\perp\,dP_\perp}{1+e^{P_x^2+P_\perp^2-\alpha}}=
$$
$$
\equiv 2\pi \int\limits_{0}^{\infty}f_F(P)P_\perp dP_\perp=
2\pi \int\limits_{0}^{\infty}\dfrac{e^{\alpha-P_x^2-P_\perp^2}P_\perp
dP_\perp}{1+e^{\alpha-P_x^2-P_\perp^2}}=$$$$=
\pi\ln(1+e^{\alpha-P_x^2}).
\eqno{(3.6)}
$$

Hence, the expression for the classical component is simplified to
one-dimensional integral
$$
\sigma_{tr}^{\rm classic}=
\dfrac{\sigma_0}{4f_2(\alpha)}
\int\limits_{-\infty}^{\infty}\dfrac{\ln(1+e^{\alpha-
P_x^2})dP_x}{1-i\omega\tau+ik_1P_x},
\eqno{(3.7)}
$$
or
$$
\sigma_{tr}^{\rm classic}=-\dfrac{\sigma_0y}{4f_2(\alpha)q}
\int\limits_{-\infty}^{\infty}\dfrac{\ln(1+e^{\alpha-\tau^2})d\tau}
{\tau-z/q},
$$
where
$$
z=\dfrac{\omega+i \nu}{k_Tv_T},  \qquad q=\dfrac{k}{k_T}.
$$

The quantum item (3.4) we present in the form of the sum of two
items
$$
\sigma_{tr}^{\rm quant}=\sigma_1+\sigma_2.
\eqno{(3.8)}
$$

Here
$$
\sigma_1=-\dfrac{\sigma_0 k_1}{4\pi f_2(\alpha)\omega\tau}\int
\dfrac{P_x(P^2-P_x^2)g(P)\,d^3P}{1-i\omega\tau +ik_1P_x},
$$
and
$$
\sigma_2=\dfrac{\sigma_0\E_T}
{4\pi f_2(\alpha)\hbar \omega}\int
\dfrac{f_F^+(\mathbf{P})-f_F^-(\mathbf{P})}{1-i\omega\tau+ik_1P_x}
(P^2-P_x^2)d^3P.
\eqno{(3.9)}
$$

With the help of the equality (3.6) the expression for $\sigma_1$
can be rewritten in the following form
$$
\sigma_1=-\dfrac{\sigma_0 k_1}{4f_2(\alpha)\omega\tau}\int
\limits_{-\infty}^{\infty}\dfrac{P_x\ln(1+e^{\alpha-
P_x^2})dP_x}{1-i\omega\tau+ik_1P_x},
\eqno{(3.10)}
$$
or
$$
\sigma_1=-\dfrac{\sigma_0 y}{4f_2(\alpha)x}\int
\limits_{-\infty}^{\infty}\dfrac{\tau\ln(1+e^{\alpha-
\tau^2})d\tau}{\tau-z/q}.
$$

After change of variable
$$
P_x\mp \dfrac{\hbar k}{2p_T}\equiv P_x\mp \dfrac{k_1\hbar \nu}
{2mv_T^2}\equiv P_x\mp \dfrac{k_1\hbar \nu}{4\E_T} \to P_x
$$
the difference of integrals from (3.9) will be transformed
to one integral and we receive \smallskip
$$
\sigma_2=-\dfrac{i\sigma_0k_1^2}{8\pi f_2(\alpha)\omega\tau}
\int
\dfrac{f_F(P)(P^2-P_x^2)\,d^3P}{(1-i\omega\tau+ik_1P_x)^2+
(k_1^2\hbar\nu/4\E_T)^2}.
\eqno{(3.11)}
$$
\medskip

In the same way, as well as during the derivation of the formula
(3.6), double internal integral in (3.11) we reduce to the
one-dimensional integral
$$
\int\limits_{-\infty}^{\infty}\int\limits_{-\infty}^{\infty}
f_F(P)[P^2-P_x^2]\;d\,\mathbf{P}_{\perp}=
\int\limits_{0}^{\infty}\int\limits_{0}^{2\pi}
f_F(P)P_{\perp}^3\,dP_{\perp}d\chi=
$$
$$
=2\pi \int\limits_{0}^{\infty}
{P}_{\perp}\ln(1+e^{\alpha-P_x^2-P_{\perp}^2})\,d{P}_{\perp}.
$$

Now the expression (3.11) can be written in the following form
(replacing a variable of integration $P_{\perp}=\rho$))
$$
\sigma_2=-\dfrac{i\sigma_0 k_1^2}{4f_2(\alpha)\omega\tau}
\int\limits_{-\infty}^{\infty}\int\limits_{0}^{\infty}
\dfrac{\rho\ln(1+e^{\alpha-\rho^2-P_x^2})d\rho\,dP_x}
{(1-i\omega\tau+ik_1P_x)^2+(k_1^2\hbar\nu/4\E_T)^2}.
\eqno{(3.12)}
$$

Hence we can to present the expression for transverse
conductivity in the form of the sum of one-dimensional (3.11) and
two-dimensional (3.12) integrals
$$
\sigma_{tr}=\dfrac{\sigma_0}{4f_2(\alpha)}
\int\limits_{-\infty}^{\infty}\dfrac{[1-(k_1/\omega\tau)P_x]\ln(1+e^{\alpha-
P_x^2})dP_x}{1-i\omega\tau+ik_1P_x}-
$$
$$
-\dfrac{i\sigma_0 k_1^2}{4 f_2(\alpha)\omega \tau}
\int\limits_{-\infty}^{\infty}\int\limits_{0}^{\infty}
\dfrac{\rho\ln(1+e^{\alpha-\rho^2-P_x^2})d\rho\,dP_x}
{(1-i\omega\tau+ik_1P_x)^2+(k_1^2\hbar\nu/4\E_T)^2}.
$$

In the expression (3.11) for $\sigma_2$ the thriple integral can be
reduced to one--dimensional integral. For this purpose in (3.11) we
pass to integration in spherical coordinates and present this
expression in the form
$$
\sigma_2=-
\dfrac{i\sigma_0k_1^2}{4f_2(\alpha)\omega\tau}
\int\limits_{0}^{\infty} f_F(P)P^4 J(P)\,dP,
$$
where
$$
J(P)=\int\limits_{-1}^{1}\dfrac{(1-\mu^2)\;d\mu}
{(1-i\omega\tau +ik_1P\mu)^2+(k_1^2\hbar\nu/4\E_T)^2}.
$$

Let's designate temporarily
$$
a=1-i\omega\tau, \qquad b=ik_1P, \qquad d=\dfrac{\hbar \nu k_1^2}{4\E_T},
$$
and rewrite the integral $J(P)$ in the form:
$$
J= \int\limits_{-1}^{1}\dfrac{(1-\mu^2)d\mu}{(a+b\mu)^2+d^2}.
$$

After change of variable $a+b\mu=t $ this integral will be rewritten
in the form
$$
J=\dfrac{1}{b^3}\int\limits_{a-b}^{a+b}\dfrac{b^2-(t-a)^2}{t^2+d^2}dt.
$$

This integral equals to
$$
J=-\dfrac{2}{b^2}+\dfrac{d^2+b^2-a^2}{b^3} \dfrac{1}{2id}
\ln\dfrac{(a+b-id)(a-b+id)}{(a+b+id)(a-b-id)}+
$$
$$
+\dfrac{a}{b^3}\ln\dfrac{(a+b-id)(a+b+id)}{(a-b-id)(a-b+id)},
$$
or
$$
J=-\dfrac{2}{b^2}+\dfrac{d^2+b^2-a^2}{2idb^3}\ln\dfrac{a^2-(b-id)^2}
{a^2-(d+id)^2}+\dfrac{a}{b^3}\ln\dfrac{(a+b)^2+d^2}
{(a-d)^2+d^2}.
$$

Considering designations for $a,b,d$, we receive
$$
J(P)\equiv J(P;\omega\tau,k_1)=$$
$$=\dfrac{2}{(k_1P)^2}
-
\dfrac{(1-i\omega\tau)^2+(k_1P)^2-
(\hbar \nu k_1^2/4\E_T)^2}{k_1^5P^3(\hbar \nu/2\E_T)}\times$$$$ \times
\ln\dfrac{(1-i\omega\tau)^2+(k_1P-\hbar \nu k_1^2/4\E_T)^2}
{(1-i\omega\tau)^2+(k_1P+\hbar \nu k_1^2/4\E_T)^2}+
$$
$$
+i\dfrac{1-i\omega\tau}{(k_1P)^3}\ln
\dfrac{(1-i\omega\tau+ik_1P)^2+(\hbar \nu k_1^2/4\E_T)^2}
{(1-i\omega\tau-ik_1P)^2+(\hbar \nu k_1^2/4\E_T)^2}.
\eqno{(3.13)}
$$

Thus, the expression of quantum transverse conductivity is defined
by one--dimensional integral
$$
\sigma_{tr}=\dfrac{\sigma_0}{4f_2(\alpha)}
\int\limits_{-\infty}^{\infty}\dfrac{1-(k_1/\omega\tau)P_x}
{1-i\omega\tau+ik_1P_x}\ln(1+e^{\alpha-P_x^2})dP_x-$$$$-
\dfrac{i\sigma_0k_1^2}{4f_2(\alpha)\omega\tau}
\int\limits_{0}^{\infty} f_F(P)P^4 J(P)\,dP,
$$
where the function $J(P)$ is defined by expression (3.13).

Let's consider the case of degenerate plasma separately.

\begin{center}
  \bf 5. DEGENERATE QUANTUM PLASMA
\end{center}

Let's return to the formula (3.1) for transverse conductivity. With
the help of (1.14)  we will
reduce it to the form
$$
\sigma_{tr}=\dfrac{2e^2m^3v_T^3}{\omega m(2\pi\hbar)^3}\int
\dfrac{\omega\tau-\mathbf{k}_1\mathbf{P}}
{1-i\omega\tau+i\mathbf{k}_1\mathbf{P}}g(P)P_\perp^2d^3P+
$$
$$
+\dfrac{e^2m^3v_T^5}{\omega(2\pi\hbar)^3\hbar}\int
\dfrac{f_F^+(\mathbf{P})-f_F^-(\mathbf{P})}{1-i\omega\tau+i
\mathbf{k}_1\mathbf{P}}P_\perp^2d^3P.
\eqno{(4.1)}
$$

In the formula (4.1) we will pass to a new dimensionless variable
$\mathbf{P}=\dfrac{\mathbf{p}}{p_F}$, where $p_F=mv_F, \; v_F $ is
the electron velocity on Fermi's surface which is supposed to be
spherical. Then we receive for $\sigma_{tr}$ the following
expression
$$
\sigma_{tr}=\dfrac{e^2m^3v_F^3}{(2\pi\hbar)^2\omega k_BT}\int
(\omega\tau-\mathbf{k}_1\mathbf{P})g(P)\dfrac{P_\perp^2\,d^3P}
{1-i\omega\tau+i\mathbf{k}_1\mathbf{P}}+
$$
$$
+\dfrac{e^2m^3v_F^5}{(2\pi\hbar)^3\omega\nu\hbar}\int
\Big[f_F^+-f_F^-\Big]\dfrac{P_\perp^2\,d^3P}
{1-i\omega\tau+i\mathbf{k}_1\mathbf{P}}.
\eqno{(4.1')}
$$

In this expression $l=v_F\tau$ is the mean free path of
electrons in degenerate plasma, $\mathbf{k}_1=\mathbf{k}l$,
$$
g(P)=\dfrac{\exp\Big(\dfrac{\E-\E_F}{k_BT}\Big)}
{\Big[1+\exp\Big(\dfrac{\E-\E_F}{k_BT}\Big)\Big]^2}=
\dfrac{\exp\Big(\dfrac{\E_F(P^2-1)}{k_BT}\Big)}{\Big[1+
\exp\Big(\dfrac{\E_F(P^2-1)}{k_BT}\Big)\Big]^2},
$$
$$
f_F^{\pm}=f_F(P_{\pm})=\dfrac{1}{1+\exp\Big[\dfrac{\E_F}{k_BT}
\Big(P_{\pm}^2-\dfrac{\mu}{\E_F}\Big)\Big]}=
\dfrac{1}{1+\exp\dfrac{\E^{\pm}-\mu}{k_BT}}.
$$

Here following designations are entered
$$
\E^{\pm}=\dfrac{1}{2m}\Big(\mathbf{p}\mp\dfrac{\hbar\mathbf{k}}{2}\Big)^2,
\quad
P_{\pm}^2=\Big(\mathbf{P}\mp \dfrac{\hbar \mathbf{k}}{2p_F}\Big)^2, \quad
\E_F=\dfrac{mv_F^2}{2}.
$$

Let's pass in $ (4.1')$ to a limit at $T\to 0$. Thus chemical potential
passes to Fermi energy of electrons on Fermi's surfaces, i.e.
$ \mu\to\E_F$. We easily will show that
$$
\lim\limits_{T\to 0}f_F^{\pm}=\Theta(\E_F-\E^{\pm})\equiv
\Theta(1-P_{\pm}^2)\equiv \Theta^{\pm},
$$
$$
\lim\limits_{T\to 0} \dfrac{g(P)}{k_BT}=-\dfrac{\partial}{\partial \E}
\Bigg[\lim\limits_{T\to 0}\dfrac{1}{1+\exp\dfrac{\E-\E_F}{k_BT}}\Bigg]=
$$
$$
=-\dfrac{\partial}{\partial\E}\Theta(\E_F-\E)=\delta(\E_F-\E).
$$

Here $\delta(x)$ is the Dirac delta-function, $\Theta(x)$ is the
Heaviside function,
$$
\Theta(x)=\left\{ \begin{array}{c}
                    1, \quad x>0, \\
                    0,\quad x<0,
                  \end{array}\right.
$$
$\E_F=\dfrac {mv_F^2}
{2}=\dfrac{p_F^2}{2m}$ is the electron kinetic energy on the Fermi surface,
$$
\E=\dfrac{mv^2}{2}=\dfrac{p^2}{2m}=\dfrac{p_x^2}{2m}+\dfrac{p_y^2}{2m}+
\dfrac{p_z^2}{2m}
$$
is the kinetic electron energy,
$$
\Theta^{\pm}(\E_F-\E^{\pm})=\left\{\begin{array}{c}
                              1,\quad \E^{\pm}<\E_F, \\
                              0,\quad \E^{\pm}>\E_F,
                            \end{array}\right.
$$
$$
\Theta^{\pm}\equiv \Theta(1-P_{\pm}^2)=\left\{\begin{array}{c}
                              1,\quad P_{\pm}<1, \\
                              0,\quad P_{\pm}>1,
                            \end{array}\right.
$$
though
$$
\E^{\pm}=\dfrac{1}{2m}\Big(p_x\mp\dfrac{\hbar k}{2}\Big)^2+
\dfrac{p_y^2}{2m}+\dfrac{p_z^2}{2m}.
$$

Hence, for transverse conductivity of degenerate quantum plasma we
have the following form
$$
\sigma_{tr}=\dfrac{e^2m^3v_F^5}{(2\pi\hbar)^3\omega}
\int
\Bigg[(\omega\tau-\mathbf{k}_1\mathbf{P})\delta(\E_F-\E)+$$$$+
\dfrac{\Theta(\E_F-\E^+)-\Theta(\E_F-\E^-)}{\hbar \nu}\Bigg]
\dfrac{P_\perp^2\,d^3P}{1-i\omega\tau+i\mathbf{k}_1\mathbf{P}}.
\eqno{(4.2)}
$$

Now with the help of the equation of state for degenerate plasma
$$
\Big(\dfrac{mv_F}{\hbar}\Big)^3=3\pi^2N^{(0)},
$$

we transform the formula (4.2) to the form
$$
\sigma_{tr}=\sigma_0\dfrac{3mv_F^2}{8\pi}\int
\Big[\big(1-\dfrac{\mathbf{k}_1\mathbf{P}}{\omega\tau}\big)\delta(\E_F-\E)
+\hspace{5cm}
$$
$$
+\dfrac{1}{\hbar\omega}\big[\Theta(\E_F-\E^+)-
\Theta(\E_F-\E^-)\big]\Big]
\dfrac{P_\perp^2\;d^3P}{1-i\omega\tau+i\mathbf{k}_1
\mathbf{P}}.
$$

Let's note, that
$$
\delta(\E_F-\E)=\delta\Big(\dfrac{mv_F^2}{2}(1-P^2)\Big)=
\dfrac{2}{mv_F^2}\delta(1-P^2)=
$$
$$
=\dfrac{1}{mv_F^2}\delta(1-P),
$$
$$
\Theta(\E_F-\E^+)=\Theta(\E_F(1-P_{\pm}^2))\equiv
\Theta(1-P_{\pm}^2).
$$

By means of this equality the expression for $\sigma_2$ can be
written in the following form
$$
\sigma_{tr}=\sigma_0\dfrac{3}{8\pi}\Bigg[\int \dfrac{\big(1-
(\omega\tau)^{-1}\mathbf{k}_1\mathbf{P}\big)\delta(1-P)}
{1-i\omega\tau+i\mathbf{k}_1\mathbf{P}}
P_\perp^2\;d^3P+\hspace{4cm}
$$
$$
+\dfrac{mv_F^2}{\hbar \omega}\int \dfrac{\Theta(1-P_+^2)-
\Theta(1-P_-^2)}{1-i\omega\tau+i\mathbf{k}_1\mathbf{P}}
P_\perp^2\;d^3P\Bigg].
\eqno{(4.3)}
$$

Here
$$
P_{\pm}^2=\Big(\mathbf{P}\mp\dfrac{\hbar
\mathbf{k}}{2p_F}\Big)^2,$$$$
\Theta(1-P_{\pm}^2)=\Theta\Big[1-\Big(\mathbf{P}\mp\dfrac{\hbar
\mathbf{k}}{2p_F}\Big)^2\Big],
$$
or,
$$\renewcommand{\arraystretch}{1.8}
\Theta^{\pm}(\mathbf{P})\equiv \Theta(1-P_{\pm}^2)=
$$\smallskip
$$=
\left\{\begin{array}{c}
1,\quad\text{if}\quad \Big(P_x\mp\dfrac{\hbar k_x}{2p_F}\Big)^2+
\Big(P_y\mp\dfrac{\hbar k_y}{2p_F}\Big)^2+
\Big(P_z\mp\dfrac{\hbar k_z}{2p_F}\Big)^2<1,\\
0,\quad\text{if}\quad \Big(P_x\mp\dfrac{\hbar k_x}{2p_F}\Big)^2+
\Big(P_y\mp\dfrac{\hbar k_y}{2p_F}\Big)^2+
\Big(P_z\mp\dfrac{\hbar k_z}{2p_F}\Big)^2>1.\end{array}\right.
$$\smallskip

Let's consider the special case of transverse conductivity, when
the wave number $k$ is equal to zero. Then in the formula (4.3) the
second (quantum) item drops out and we obtain
$$
\sigma_{tr}(k=0)=\sigma_{tr}^{\rm classic}(k=0)=$$$$=
\sigma_0\dfrac{3}{8\pi}\dfrac{\nu}{\nu-i\omega}\int
\delta(1-P)[P^2-P_x^2]\,d^3P,
$$
whence we receive the well known formula for classical plasma
$$
\sigma_{tr}(k=0)=\sigma_{tr}^{\rm classic}(k=0)=
\sigma_0\dfrac{\nu}{\nu-i\omega}.
$$

Further everywhere we will direct the axis $x$ along the vector
$\mathbf{k}_1$. Let's consider a case of small values of the
product $\hbar k$.
We will note that
$$\renewcommand{\arraystretch}{1.8}
\Theta^{\pm}\equiv \Theta(\E_F-\E^{\pm})=\left\{
      \begin{array}{c}
        1,\quad \text{if}\quad
        \Big(p_x\mp\dfrac{\hbar k}{2}\Big)^2+p_y^2+p_z^2<p_F^2, \\
        0,\quad \text{if}\quad
        \Big(p_x\mp\dfrac{\hbar k}{2}\Big)^2+p_y^2+p_z^2>p_F^2 \\
      \end{array}.
    \right.
$$

Here
$$
\E^{\pm}=\dfrac{1}{2m}\Big(p_x\mp\dfrac{\hbar k}{2}\Big)^2+
\dfrac{1}{2m}p_y^2+\dfrac{1}{2m}p_z^2.
$$

Let's expand $\Theta^{\pm}$ by powers of $\hbar k$ to the second
order inclusive
$$
\Theta^{\pm}=\Theta(\E_F-\E^{\pm})=\Theta(1-P_{\pm}^2)=$$$$=
\Theta(1-P^2)\pm\delta(1-P^2)P_x\dfrac{\hbar k}{p_F}+$$$$+
\Big[\delta'(1-P^2)P_x^2-\dfrac{1}{2}\delta(1-P^2)\Big]
\dfrac{\hbar^2k^2}{2p_F^2}\pm
$$
$$
\pm \Big[\delta''(1-P^2)P_x^3-\dfrac{3}{2}\delta'(1-P^2)P_x\Big]
\dfrac{\hbar^3k^3}{6p_F^3}.
$$

From here follows that
$$
\Theta^+(1-P_+^2)-\Theta^-(1-P_-^2)=
2\delta(1-P^2)P_x\dfrac{\hbar k}{p_F}+$$$$+
\Big[\delta''(1-P^2)P_x^3-\dfrac{3}{2}\delta'(1-P^2)P_x\Big]
\dfrac{\hbar^3k^3}{3p_F^3}.
$$

We will present the formula for calculation of transverse conductivity in
the following form
$$
\sigma_{tr}=\sigma_0 f_{tr},
$$
where
$$
f_{tr}=\dfrac{3}{8\pi\omega\tau}
\int \Big[(\omega\tau-k_1P_x)\delta(1-P^2)+ \hspace{4cm}
$$
$$ \hspace{3cm}
+\dfrac{mv_F^2}{\hbar \nu}(\Theta^+-\Theta^-)\Big]
\dfrac{(P^2-P_x^2)\;d^3P}{1-i\omega\tau+ik_1P_x},
$$
or
$$
f_{tr}=\dfrac{3}{8\pi}\int \Big[\Big(1-\dfrac{k_1P_x}{\omega\tau}\Big)
\delta(1-P)+\dfrac{mv_F^2}{\hbar \omega}(\Theta^+-\Theta^-)\Big]
\dfrac{(P^2-P_x^2)\;d^3P}{1-i\omega\tau+ik_1P_x}.
$$

Let's consider the integrand from the equality for $f_{tr}$
$$
\Big(1-\dfrac{k_1P_x}{\omega\tau}\Big)
\delta(1-P)+\dfrac{mv_F^2}{\hbar \omega}(\Theta^+-\Theta^-)=
$$
$$
=\delta(1-P)-\dfrac{k_1P_x}{\omega\tau}\delta(1-P)+
2\delta(1-P^2)P_x\dfrac{mv_F^2k}{\omega p_F}+
$$
$$
+\dfrac{mv_F^2}{\hbar \omega}
\Big[\delta''(1-P^2)P_x^3-\dfrac{3}{2}\delta'(1-P^2)P_x\Big]
\dfrac{\hbar^3k^3}{3p_F^3}=
$$
$$
=\delta(1-P)+\Big[\delta''(1-P^2)P_x^3-\dfrac{3}{2}\delta'(1-P^2)P_x\Big]
\Big(\dfrac{\hbar \nu}{mv_F^2}\Big)^2\dfrac{k_1^3}{3\omega\tau}.
$$

Therefore the transverse conductivity at small $\hbar k$ is equal
to
$$
\sigma_{tr}=\sigma_{tr}^{\rm classic}+$$$$+\dfrac{\hbar^2 \nu^2k_1^3}
{8\pi(mv_F^2)^2\omega\tau}\int \dfrac{\delta''(1-P^2)P_x^2-\frac{3}{2}
\delta'(1-P^2)}{1-i\omega\tau+ik_1P_x}P_x(P^2-P_x^2)\,d^3P.
$$ \medskip

Here $k_1=kl,\;l=v_F\tau$ is the electron mean free path,
$$
\sigma_{tr}^{\rm classic}=\sigma_0\dfrac{3}{8\pi}\int
\dfrac{\delta(1-P)(P^2-P_x^2)\,d^3P}{1-i\omega\tau+ik_1P_x}.
$$

From the first formula follows that under $\hbar\to 0\quad
\sigma_{tr}\to \sigma_{tr}^{\rm classic}$, i.e. under tendency of
Planck's constant to zero the transverse conductivity passes into
the classical.

Let's pass to decomposition on degrees of wave number of the quantum
component the transverse conductivity.
For this purpose we will spread out by degrees
of wave number of Fermi --- Dirac distributions 
$\Theta^{\pm}=\Theta(1-P_{\pm}^2)$. We obtain that
$$
\Theta^{\pm}=\Theta(1-P_{\pm}^2)=\Theta(1-P^2)\pm
\delta(1-P^2)P_x\dfrac{\hbar k}{2p_F}+$$$$+\Big[\delta'(1-P^2)P_x^2-
\dfrac{1}{2}\delta(1-P^2)\Big]\dfrac{\hbar^2 k^2}{2p_F^2}\pm
$$
$$
\pm\Big[\delta''(1-P^2)P_x^3-\dfrac{3}{2}\delta'(1-P^2)P_x\Big]
\dfrac{h^3k^3}{6p_F^3}+
$$
$$
+\Big[\delta'''(1-P^2)P_x^4-3\delta''(1-P^2)P_x^2+\dfrac{3}{4}
\delta'(1-P^2)\Big]\dfrac{\hbar^4k^4}{24p_F^4}\pm
$$
$$
\pm\Big[\delta^{(4)}(1-P^2)P_x^5-5\delta'''(1-P^2)P_x^3+
\dfrac{15}{4}\delta''(1-P^2)P_x\Big]\dfrac{\hbar^5k^5}{120p_F^5}+\cdots.
$$

From these decomposition we receive their difference
$$
\Theta^+-\Theta^-=2\delta(1-P^2)P_x\dfrac{\hbar k}{p_F}+
$$
$$
+\Big[\delta''(1-P^2)P_x^3-\dfrac{3}{2}\delta'(1-P^2)P_x\Big]
\dfrac{\hbar^3k^3}{3p_F^3}+
$$
$$
+\Big[\delta^{(4)}(1-P^2)P_x^5-5\delta'''(1-P^2)P_x^3+
\dfrac{15}{4}\delta''(1-P^2)P_x\Big]\dfrac{\hbar^5k^5}{60p_F^5}+\cdots.
$$

Now decomposition of quantum component transverse conductivity
has following expansion by degrees of wave number
$$
\sigma_{tr}^{\rm quant}=
i\sigma_0\dfrac{7(\omega\tau+i)^2k_1^6+10k_1^8}
{140\omega\tau(\omega\tau+i)^6}\Big(\dfrac{\hbar
\nu}{2\E_F}\Big)^2+\cdots,
$$
or
$$
\sigma_{tr}^{\rm quant}=i\sigma_0
\Big[\dfrac{\nu v_F^4}{20\omega m^2(\omega+i \nu)^4}(\hbar^2
k^6)+\dfrac{\nu v_F^4}{14\omega m^2(\omega+i \nu)^6}(\hbar^2 k^8)+
\cdots\Big].
$$

The expression (4.3) for transverse conductivity we will present as
the sum of two terms
$$
\sigma_{tr}=\sigma_{tr}^{\rm classic}+\sigma_{tr}^{\rm quant}.
\eqno{(4.4)}
$$

In the equality (4.4) the following designations are entered
$$
\sigma_{tr}^{\rm classic}=\sigma_0\dfrac{3}{8\pi}\int
\dfrac{\delta(1-P)(P^2-P_x^2)\,d^3P}{1-i\omega\tau+ik_1P_x},
\eqno{(4.5)}
$$
$$
\sigma_{tr}^{\rm quant}=\sigma_0\dfrac{3\nu}{8\pi\omega}
\Bigg[-k_1\int
\dfrac{P_x\delta(1-P)(P^2-P_x^2)\,d^3P}{1-i\omega \tau+ik_1P_x}+\hspace{3cm}
$$
$$
\hspace{3cm}+
\dfrac{mv_F^2}{\hbar \nu}\int
\dfrac{\Theta^+(\mathbf{P})-\Theta^-(\mathbf{P})}
{1-i\omega \tau+ik_1P_x}(P^2-P_x^2)d^3P\Bigg].
\eqno{(4.6)}
$$

The expression (4.5) for $\sigma_{tr}^{\rm classic}$ is easily
calculated in the explicit form
$$
\sigma_{tr}^{\rm classic}=-\sigma_0\dfrac{3}{4}\Big[2\nu\dfrac{\nu-i\omega}
{k^2v_F^2}-i \nu\dfrac{(\nu-i\omega)^2+k^2v_F^2}{k^3v_F^3}\ln
\dfrac{\nu-i\omega+ikv_F}{\nu-i\omega-ikv_F}\Big],
\eqno{(4.7)}
$$
or, in dimensionless parameters,
$$
\sigma_{tr}^{\rm classic}=-\sigma_0\dfrac{3}{4}
\Big[2\dfrac{1-i\omega\tau}{k_1^2}+i\dfrac{(1-i\omega\tau)^2+k_1^2}
{k_1^3}\ln\dfrac{1-i\omega \tau+ik_1}{1-i\omega\tau-ik_1}\Big].
$$

For the previous formula (4.7) we can give also such forms
$$
\sigma_{tr}^{\rm classic}=\sigma_0\dfrac{3i}{4}
\Big[2\dfrac{\omega\tau+i}{k_1^2}+\dfrac{(\omega\tau+i)^2-k_1^2}
{k_1^3}\ln\dfrac{\omega \tau+i-k_1}{\omega\tau+i+k_1}\Big],
$$
and
$$
\sigma_{tr}^{\rm classic}=\sigma_0\dfrac{3i}{4}
\Big[2\nu\dfrac{\omega+i \nu}
{k^2v_F^2}+ \nu\dfrac{(\omega+i \nu)^2-k^2v_F^2}{k^3v_F^3}\ln
\dfrac{\omega+i \nu-kv_F}{\omega+i \nu+ikv_F}\Big],
$$

We introduce the dimensionless variables
$$
z=\dfrac{\omega+i \nu}{k_Fv_F}=x+iy, \quad
x=\dfrac{\omega}{k_Fv_F}, \quad y=\dfrac{\nu}{k_Fv_F}, \quad
q=\dfrac{k}{k_F},
$$
where $k_F$ is the Fermi wave number,
$k_F=\dfrac{mv_F}{\hbar}=\dfrac{p_F}{\hbar}$.

Then we can rewrite the formula (4.7) in the form
$$
\sigma_{tr}^{\rm classic}=\sigma_0\dfrac{3i}{4}\Big[2\dfrac{yz}{q^2}+
y\dfrac{z^2-q^2}{q^3}\ln\dfrac{z-q}{z+q}\Big].
\eqno{(4.8)}
$$

We present the formula (4.6) in the form of the sum of two
components
$$
\sigma_{tr}^{\rm quant}=\sigma_1+\sigma_2.
\eqno{(4.9)}
$$

Here
$$
\sigma_1=-\sigma_0\dfrac{3}{4}\dfrac{\nu v_Fk}{\omega}J_1=
-\sigma_0\dfrac{3}{4}\dfrac{\nu}{\omega}k_1J_1,
$$
where
$$
J_1=\dfrac{1}{2\pi}\int \dfrac{P_x\delta(1-P)(P^2-P_x^2)}
{1-i\omega\tau+ik_1P_x}d^3P,
\eqno{(4.10)}
$$
and
$$
\sigma_2=\sigma_0\dfrac{3}{4\pi}\dfrac{\nu}{\omega}
\dfrac{mv_F^2}{\hbar}J_2,
$$
where
$$
J_2=\dfrac{1}{2}\int
\dfrac{\Theta^+(\mathbf{P})-\Theta^-(\mathbf{P})}
{\nu-i\omega+ikv_FP_x}(P^2-P_x^2)d^3P=
$$
$$
=\dfrac{1}{2}\int
\dfrac{\Theta^+(\mathbf{P})-\Theta^-(\mathbf{P})}
{\nu-i\omega+ikv_FP_x}(P_y^2+P_z^2)d^3P=
$$
$$
=\int
\dfrac{\Theta^+(\mathbf{P})-\Theta^-(\mathbf{P})}
{\nu-i\omega+ikv_FP_x}P_y^2\;d^3P.
\eqno{(4.11)}
$$

Calculating the integral $J_1$ in spherical coordinates, we receive,
that
$$
J_1=\int\limits_{-1}^{1}\dfrac{\mu(1-\mu^2)d\mu}{1-i\omega\tau+ik_1\mu}=
$$
$$
=-\dfrac{4i}{3k_1}\Bigg[1+\dfrac{3(1-i\omega\tau)^2}{2k_1^2}+
\dfrac{3i(1-i\omega\tau)}{4k_1^3}[(1-i\omega\tau)^2+k_1^2]
\ln\dfrac{1-i\omega\tau+ik_1}{1-i\omega\tau-ik_1}\Bigg].
$$

Therefore, the quantity $\sigma_1$ is equal to
$$
\sigma_1=\dfrac{i\sigma_0}{\omega\tau}
\Bigg[1+\dfrac{3(1-i\omega\tau)^2}{2k_1^2}+\hspace{9cm}
$$
$$
+\dfrac{3i(1-i\omega\tau)}{4k_1^3}[(1-i\omega\tau)^2+k_1^2]
\ln\dfrac{1-i\omega\tau+ik_1}{1-i\omega\tau-ik_1}\Bigg],
\eqno{(4.12)}
$$
or
$$
\sigma_1=i\sigma_0\dfrac{\nu}{\omega}\Bigg[1-\dfrac{3(\omega+i \nu)^2}
{2k^2v_F^2}-\dfrac{3(\omega+i \nu)}{k^3v_F^3}\Big[(\omega+i
\nu)^2-$$$$\hspace{5cm}-
k^2v_F^2\Big]\ln\dfrac{\omega+i \nu-kl}{\omega+i \nu+kl}\Bigg].
$$

Let's pass to calculation of the summand $\sigma_2$ which we will
present in the form
$$
\sigma_2=\sigma_0\dfrac{3\nu}{4\omega}\dfrac{mv_F^2}{\pi\hbar \nu}
\int \dfrac{\Theta^+(\mathbf{P})-\Theta^-(\mathbf{P})}{1-i\omega\tau+
ik_1P_x}P_y^2\,d^3P.
\eqno{(4.13)}
$$

Let's present the formula (4.13) in the form of the difference
$$
\sigma_2=\sigma_0\dfrac{3\nu}{4\omega}\dfrac{mv_F^2}{\pi\hbar \nu}
\Bigg[\int \dfrac{\Theta^+(\mathbf{P})P_y^2\,d^3P}{1-i\omega\tau+
ik_1P_x}-
\int \dfrac{\Theta^-(\mathbf{P})P_y^2\,d^3P}{1-i\omega\tau+
ik_1P_x}\Bigg],
$$
or
$$
\sigma_2=\sigma_0\dfrac{3\nu}{4\omega}\dfrac{mv_F^2}{\pi\hbar \nu}
(J^+-J^-),
$$
where
$$
J^{\pm}=\int
\dfrac{\Theta^{\pm}(\mathbf{P})P_y^2\,d^3P}{1-i\omega\tau+
ik_1P_x}.
$$

It is obvious that these integrals are equal to:
$$
J^{\pm}=\int\limits_{S^{\pm}_3}
\dfrac{P_y^2\,d^3P}{1-i\omega\tau+ik_1P_x}.
$$

Here $S^{\pm}_3=S_3\Big(\pm\dfrac{\hbar k}{2p_F},0,0\Big)$ is
the three-dimensional sphere with unitary radius with the centre in
the point $\Big(\pm\dfrac{\hbar k}{2p_F},0,0\Big)$
$$
S^{\pm}_3\Big(\pm\dfrac{\hbar k}{2p_F},0,0\Big)\equiv
\left\{(P_x,P_y,P_z):\Big(P_x\mp\dfrac{\hbar
k}{2p_F}\Big)^2+P_y^2+P_z^2<1\right\}.
$$

After obvious replacement of variables we receive, that
$$
J^{\pm}=\int\limits_{S_3^{\pm}}
\dfrac{P_y^2\;d^3P}{1-i\omega\tau+ik_1P_x}=
\int\limits_{S_3(\mathbf{0})}\dfrac{P_y^2\;d^3P}{1-i\omega\tau+ik_1
\Big(P_x\pm\dfrac{\hbar\nu k_1}{2mv_F^2}\Big)},
$$
where $S_3(\mathbf{0})$ is the sphere with the centre in zero with
unitary radius,
$$
S_3(\mathbf{0})=\left\{(P_x,P_y,P_z):P_x^2+P_y^2+P_z^2<1\right\}.
$$

Now it is easy to find, that the item $\sigma_2$ is calculated by
the formula
$$
\sigma_2=-i\sigma_0\dfrac{3k_1^2}{8\pi
\omega\tau}\int\limits_{S_3(\mathbf{0})}\dfrac{(P_y^2+P_z^2)\,d^3P}
{(1-i\omega\tau+ik_1P_x)^2+\Big(\dfrac{\hbar\nu k_1^2}{2mv_F^2}\Big)^2}.
\eqno{(4.14)}
$$

We present the sphere $S_3(\mathbf{0})$ in the form of joining:
$$
S_3(\mathbf{0})=\bigcup\limits_{P_x=-1}^{P_x=1}S^2_{1-P_x^2}(0,0).
$$

Here $S^2_{1-P_x^2}(0,0)$ is the circle of the form:
$$
S^2_{1-P_x^2}(0,0)=\left\{(P_y,P_z): P_y^2+P_z^2<1-P_x^2 \right\}.
$$

Now we will calculate the integrals $J^{\pm}$ as repeated
$$
J^{\pm}=\dfrac{1}{2}
\int\limits_{-1}^{1}
\dfrac{dP_x}{1-i\omega\tau+ik_1 \Big(P_x\pm
\dfrac{\hbar\nu k}{2mv_F^2}\Big)}\;
\iint\limits_{S^2_{1-P_x^2}(0,0)}(P_y^2+P_z^2)\;d \mathbf{P_{\perp}}=
$$
$$
=\dfrac{1}{2}
\int\limits_{-1}^{1}
\dfrac{dP_x}{1-i\omega\tau+ik_1 \Big(P_x\pm
\dfrac{\hbar \nu k_1}{2mv_F^2}\Big)}\;
\int\limits_{0}^{1-P_x^2}\int\limits_{0}^{2\pi}
P_{\perp}^3\,dP_{\perp}\,d\chi=
$$
$$
=\dfrac{\pi}{4}\int\limits_{-1}^{1}
\dfrac{(1-t^2)^2dt}
{1-i\omega \tau+ik_1 \Big(t\pm \dfrac{\hbar \nu k_1}{2mv_F^2}\Big)}, \qquad
t=P_x.
$$

We note that
$$
ik_1\dfrac{\hbar\nu k_1}{2mv_F^2}=
i\dfrac{\hbar \nu k_1^2}{2mv_F^2}=i\dfrac{\hbar \nu k_1^2}
{4\E_F}=ic_0,
$$
moreover
$$
c_0=\dfrac{\hbar \nu k_1^2}{2mv_F^2}=\dfrac{k_1^2}{2k_Fl}=
\dfrac{k_1^2}{2k_{01}}.
$$

Here $k_{01}$ is the dimensionless Fermi wave number, $
k_{01}=k_Fl. $

Now the summand $\sigma_2$  is equal to
$$
\sigma_2=\sigma_0\dfrac{3\nu}{4\omega}\dfrac{mv_F^2}{\pi\hbar \nu}
[J^+-J^-]=
$$
$$
=\sigma_0\dfrac{3\nu}{16\omega}\dfrac{mv_F^2}{\hbar \nu}
\int\limits_{-1}^{1}\Big[\dfrac{(1-t^2)^2}{1-i\omega\tau+ik_1t+ic_0}-
\dfrac{(1-t^2)^2}{1-i\omega\tau+ik_1t-ic_0}\Big]dt=
$$
$$
=-i\sigma_0\dfrac{3 k_1^2}{16\omega\tau}\int\limits_{-1}^{1}
\dfrac{(1-t^2)^2\,dt}{(1-i\omega\tau+ik_1t)^2+c_0^2}.
$$

We consider the denominator
$$
(1-i\omega\tau+ik_1t)^2+c_0^2=-\Big[k_1^2
\Big(t-\dfrac{\omega\tau+i}{k_1}\Big)^2-c_0^2\Big]=
$$
$$
=-k_1^2\Big[\Big(t-\dfrac{\omega\tau+i}{k_1}\Big)^2-
\Big(\dfrac{k_1\hbar}{2p_Fl}\Big)^2\Big]=-k_1^2\Big[(t-a)^2-c^2\Big],
$$
where
$$
a=\dfrac{\omega\tau+i}{k_1}=\dfrac{z}{q}, \qquad
c=\dfrac{k_1\hbar}{2p_Fl}=
\dfrac{k_1}{2k_Fl}=\dfrac{k_1}{2k_{01}}=\dfrac{q}{2}.
$$

Thus, the expression (4.14) can be written in the form
$$
\sigma_2=i\sigma_0\dfrac{3}{16\omega\tau} J,
\eqno{(4.15)}
$$
where
$$
J=\int\limits_{-1}^{1}\dfrac{(1-t^2)^2}{(t-a)^2-c^2}.
$$

After variable replacement $t-a=x, \; x_0 =-1-a, \; x_1 =-1+a $,
we receive, that
$$
J=\int\limits_{x_0}^{x_1}\dfrac{[1-(x-a)^2]^2}{x^2-c^2}dx=
$$
$$
=\dfrac{(a^2+c^2-1)^2+4a^2c^2}{2c}\ln\dfrac{a^2-(c-1)^2}{a^2-(c+1)^2}+
6a^2+2c^2-\dfrac{10}{3}+
$$
$$
+2a(a^2+c^2-1)\ln\dfrac{(a-1)^2-c^2}{(a+1)^2-c^2}.
$$

The value of this integral in parameters $z $ and $q $ is expressed
by the equality
$$
J=\dfrac{(z^2-q^2+q^4/4)^2+z^2q^4}{q^5}\ln\dfrac{z^2-(q-q^2/2)^2}
{z^2-(q+q^2/2)^2}+6\dfrac{z^2}{q^2}+\dfrac{q^2}{2}-\dfrac{10}{3}+
$$
$$
+2\dfrac{z}{q^3}(z^2-q^2+q^4/4)\ln\dfrac{(z-q)^2-q^4/4}{(z+q)^2-q^4/4}.
$$

The expression (4.15) for $\sigma_2$ with the help of the previous
expression for $J$ we present in the form
$$
\sigma_2=i\sigma_0\dfrac{3y}{8x}\Bigg[-\dfrac{5}{3}+3\dfrac{z^2}{q^2}+
\dfrac{q^2}{4}+\dfrac{1}{2q^5}\Big[\Big(z^2-q^2+\dfrac{q^4}{4}\Big)^2+
z^2q^4\Big]\times
$$
$$
\times\ln\dfrac{z^2-(q-q^2/2)^2}{z^2-(q+q^2/2)}+
\dfrac{z}{q^2}\Big(z^2-q^2+\dfrac{q^4}{4}\Big)\ln\dfrac{(z-q)^2-q^4/4}
{(z+q)^2-q^4/4}\Bigg].
\eqno{(4.16)}
$$

We will present the formula (4.12) for $\sigma_1$ in terms of $z$ and
$q$
$$
\sigma_1=i\sigma_0\dfrac{y}{x}\Big[1-\dfrac{3z^2}{2q^2}-
\dfrac{3z}{4q^3}(z^2-q^2)\ln\dfrac{z-q}{z+q}\Big],
\eqno{(4.17)}
$$

With the help of the equalities (4.16) and (4.17) we get the quantum
part of transverse permittivity
$$
\sigma_{tr}^{\rm quant}=i\sigma_0\dfrac{3y}{8x}\Bigg[1-\dfrac{z^2}{q^2}
+\dfrac{q^2}{4}-\dfrac{2z}{q^3}(z^2-q^2)\ln\dfrac{z-q}{z+q}+\hspace{4cm}
$$
$$
+\dfrac{1}{2q^5}\Big[\Big(z^2-q^2+q^4/4\Big)^2+z^2q^4\Big]
\ln\dfrac{z^2-(q-q^2/2)^2}{z^2-(q+q^2/2)^2}+
$$
$$
+\dfrac{z}{q^3}\Big(z^2-q^2+\dfrac{q^4}{4}\Big)
\ln\dfrac{(z-q)^2-q^4/4}{(z+q)^2-q^4/4}\Bigg].
\eqno{(4.18)}
$$

Let's note, that the expressions (4.17) and (4.18) contain Kohn
singularities in the form $x\ln x$, where $x=z-q$, or $x=z+q$.

Now it is necessary to sum the quantum term (4.18) and classical
term (4.8), for this purpose we present the expression (4.8) in the
similar to (4.18) form
$$
\sigma_{tr}^{\rm classic}=i\sigma_0\dfrac{3y}{8x}\Bigg[\dfrac{4xz}{q^2}+
\dfrac{2x}{q^3}(z^2-q^2)\ln\dfrac{z-q}{z+q}\Bigg].
\eqno{(4.19)}
$$

Adding (4.18) and (4.19), we receive the final expression for the
transverse permittivity in quantum plasma
$$
\sigma_{tr}(x,y,q)=i\sigma_0\dfrac{3y}{8x}\Bigg[1+
\dfrac{z(3x-iy)}{q^2}+\dfrac{q^2}{4}-
\dfrac{2iy}{q^3}(z^2-q^2)\ln\dfrac{z-q}{z+q}+\hspace{2cm}
$$
$$
+\dfrac{1}{2q^5}\Big[\Big(z^2-q^2+q^4/4\Big)^2+z^2q^4\Big]
\ln\dfrac{z^2-(q-q^2/2)^2}{z^2-(q+q^2/2)^2}+
$$
$$
+\dfrac{z}{q^3}\Big(z^2-q^2+\dfrac{q^4}{4}\Big)
\ln\dfrac{(z-q)^2-q^4/4}{(z+q)^2-q^4/4}\Bigg].
\eqno{(4.20)}
$$ \smallskip

It is more convenient for graphic research of conductivity instead of
formulas (4.20) to use the equivalent form
$$
\dfrac{\sigma_{tr}}{\sigma_0}=-iy\dfrac{3}{4}\int\limits_{-1}^{1}
\dfrac{(1-t^2)dt}{qt-z}+i\dfrac{3yq}{4x}\int\limits_{-1}^{1}
\dfrac{t(1-t^2)dt}{qt-z}+
$$
$$
+i\dfrac{3yq^2}{16x}
\int\limits_{-1}^{1}\dfrac{(1-t^2)^2dt}{(qt-z)^2-q^4/4}.
\eqno{(4.21)}
$$

\begin{center}\bf
  6. COMPARISON WITH LINDHARD'S FORMULAS
\end{center}

Let's consider Lindhard formula (5.3.4) from \cite {Dressel} for
transverse conductivity and let's transform  it to our
designations. After limiting transition from this formula (5.3.4)
we receive
$$
\hat{\sigma}(\mathbf{q},\omega)=\dfrac{i\sigma_0}{\omega\tau}-
\dfrac{i}{\Omega}\dfrac{e^2}{\omega m^2}\sum\limits_{\mathbf{k}}\Bigg[
\dfrac{f^0(\E_k)}{\E_{\mathbf{k}}-\E_{\mathbf{k-q}}-
\hbar(\omega-i\nu)}-\hspace{3cm}
$$
$$
\hspace{5cm}-
\dfrac{f^0(\E_{k})}{\E_{\mathbf{k+q}}-\E_{\mathbf{k}}-
\hbar(\omega-i\nu)}\Bigg]|\langle\mathbf{k}+\mathbf{q}|\mathbf{p}|
\mathbf{k}\rangle_*|^2.
\eqno{(5.1)}
$$

Here $f^0(\E_k)$ is the absolute Fermi --- Dirac distribution,
$\E_k=\dfrac{\hbar^2k^2}{2m}$; besides, the sum from (5.1) is to be
understood as integral
$$
\dfrac{1}{\Omega}\sum\limits_{\mathbf{k}}=
\int \dfrac{2d\mathbf{k}}{(2\pi)^3}.
$$

Let's present the formula (5.1) in the form
$$
\hat{\sigma}(\mathbf{q},\omega)=\dfrac{i\sigma_0}{\omega\tau}+
\hat{\sigma}_2,
\eqno{(5.2)}
$$
where
$$
\hat{\sigma}_2=-
\dfrac{i}{\Omega}\dfrac{e^2}{\omega m^2}\sum\limits_{\mathbf{k}}\Bigg[
\dfrac{f^0(\E_k)}{\E_{\mathbf{k}}-\E_{\mathbf{k-q}}-
\hbar(\omega-i\nu)}-\hspace{4cm}
$$
$$
\hspace{5cm}-
\dfrac{f^0(\E_{k})}{\E_{\mathbf{k+q}}-\E_{\mathbf{k}}-
\hbar(\omega-i\nu)}\Bigg]|\langle\mathbf{k}+\mathbf{q}|\mathbf{p}|
\mathbf{k}\rangle_*|^2.
\eqno{(5.3)}
$$

Let's present the formula (5.3) in the integrated form
$$
\hat{\sigma}_2=-\dfrac{ie^2}{m\omega}\int
\Theta(\E_F-\E)\dfrac{2d\mathbf{k}}{(2\pi)^3}
\Bigg[k^2-\Big(\dfrac{\mathbf{k}\mathbf{q}}{q}\Big)^2\Bigg]
\times $$$$ \times
\Bigg[\dfrac{1}{q^2+2\mathbf{k}\mathbf{q}-\dfrac{2m}{\hbar}(\omega+
i/\tau)}+\dfrac{1}{q^2-2\mathbf{k}\mathbf{q}+\dfrac{2m}{\hbar}(\omega+
i/\tau)}\Bigg].
\eqno{(5.4)}
$$

Let's transform the formula (5.4). We will direct the wave vector
$\mathbf{q}$ along the  $x$--component of momentum, i.e. we
take $\mathbf{q}=\{k,0,0\} $, and instead of the vector
$\mathbf{k}$ we enter the dimensionless vector $\mathbf{P} $ with
unit length by the following equality
$$
\mathbf{k}=\dfrac{\mathbf{p}}{\hbar}=\dfrac{p_F}{\hbar}\mathbf{P},
\qquad p_F=mv_F.
$$

Therefore
$$
k^2-\Big(\dfrac{\mathbf{k}\mathbf{q}}{q}\Big)^2=
\dfrac{p_F^2}{\hbar^2}\Big(P^2-P_x^2\Big)
=\dfrac{p_F^2}{\hbar^2}\Big(P_y^2+P_z^2\Big),
$$\medskip
$$
\dfrac{2d\mathbf{k}}{(2\pi)^3}=\dfrac{2d^3p}{(2\pi\hbar)^3}
=\dfrac{2p_F^3\,d^3P}{(2\pi\hbar)^3}\equiv d\Omega_F.
$$

The  absolute Fermi --- Dirac distribution in our designations has
the following form
$$f^0(\E_k)=\Theta(\E_F-\E),\qquad
\E=\dfrac{p^2}{2m}=\dfrac{p_F^2}{2m}P^2=\E_FP^2, \quad
\E_F=\dfrac{p_F^2}{2m}.
$$

Noticing, that the absolute  Fermi --- Dirac distribution is
normalized in terms of numerical density, i.e.
$$
\int \Theta(\E_F-\E)d\Omega_F=N,
$$
we transform the second  square brackets from (5.4). We have
$$
\dfrac{1}{q^2+2\mathbf{k}\mathbf{q}-\dfrac{2m}{\hbar}(\omega+
i/\tau)}+\dfrac{1}{q^2-2\mathbf{k}\mathbf{q}+\dfrac{2m}{\hbar}(\omega+
i/\tau)}=
$$
$$
=\dfrac{i\hbar}{2m}\Bigg[\dfrac{1}{\nu-i\omega+ik v_FP_x+
i\dfrac{\hbar k^2}{2m}}-
\dfrac{1}{\nu-i\omega+ik v_FP_x-i\dfrac{\hbar k^2}{2m}}\Bigg]=
$$
$$
=\dfrac{\hbar^2 k^2}{2m^2}\dfrac{1}{(\nu-i \omega+ikv_FP_x)^2+
\Big(\dfrac{\hbar k^2}{2m}\Big)^2}=
$$
$$
=\dfrac{\hbar^2k_1^2}{2p_F^2}
\dfrac{1}{(1-i\omega\tau+ik_1P_x)^2+
\Big(\dfrac{\hbar\nu k_1^2}{2mv_F^2}\Big)^2}.
$$

Now we receive the integrated summand of Lindhard in the form
$$
\hat{\sigma}_2=-
\dfrac{i\sigma_03k_1^2}{8\pi\omega\tau}\int
\dfrac{\Theta(\E_F-\E)
(P^2-P_x^2)\,d^3P}{(1-i\omega\tau+ik_1P_x)^2+
\Big(\dfrac{\hbar \nu k_1^2}{2m v_F^2}\Big)^2},
$$
or, that is the same that
$$
\hat{\sigma}_2=-
\dfrac{i\sigma_03k_1^2}{8\pi\omega\tau}\int\limits_{S_3(\mathbf{0})}
\dfrac{(P_y^2+P_z^2)\,d^3P}{(1-i\omega\tau+ik_1P_x)^2+
\Big(\dfrac{\hbar \nu k_1^2}{2m v_F^2}\Big)^2},
\eqno{(5.5)}
$$

The formula (5.5) precisely coincides with the formula (4.14) for
$\sigma_2$.  The $\sigma_2$ is calculated according to (4.16)
$$
\sigma_2=i\sigma_0\dfrac{3y}{8x}\Bigg[-\dfrac{5}{3}+3\dfrac{z^2}{q^2}+
\dfrac{q^2}{4}+\dfrac{1}{2q^5}\Big[\Big(z^2-q^2+\dfrac{q^4}{4}\Big)^2+
z^2q^4\Big]\times
$$
$$
\times\ln\dfrac{z^2-(q-q^2/2)^2}{z^2-(q+q^2/2)}+
\dfrac{z}{q^3}\Big(z^2-q^2+\dfrac{q^4}{4}\Big)\ln\dfrac{(z-q)^2-q^4/4}
{(z+q)^2-q^4/4}\Bigg].
\eqno{(5.6)}
$$

In the monograph \cite {Dressel} the following formula
(the formula (5.3.6) from \cite {Dressel}) for calculation of
conductivity is presented
$$
\hat{\sigma}(\mathbf{q},\omega)=\dfrac{iNe^2}{\omega m}
\Bigg(\dfrac{3}{8}\Big[\Big(\dfrac{q}{2k_F}\Big)^2+3\Big(
\dfrac{\omega+i/\tau}{qv_F}\Big)^2+1\Big]-\hspace{3cm}
$$
$$
-\dfrac{3k_F}{16q}\Bigg[1-\Big(\dfrac{q}{2k_F}-\dfrac{\omega+i/\tau}
{qv_F}\Big)^2\Bigg]^2{\rm Ln}\left\{\dfrac{\dfrac{q}{2k_F}-
\dfrac{\omega+i\ \tau}{qv_F}+1}{\dfrac{q}{2k_F}-
\dfrac{\omega+i\ \tau}{qv_F}-1}\right\}-
$$
$$
-\dfrac{3k_F}{16q}\Bigg[1-\Big(\dfrac{q}{2k_F}+\dfrac{\omega+i/\tau}
{qv_F}\Big)^2\Bigg]^2{\rm Ln}\left\{\dfrac{\dfrac{q}{2k_F}+
\dfrac{\omega+i\ \tau}{qv_F}+1}{\dfrac{q}{2k_F}+
\dfrac{\omega+i\ \tau}{qv_F}-1}\right\}\Bigg).
\eqno{(5.7)}
$$

Now we can rewrite the formula (5.7) with the use of our notations
$$
\sigma_{tr}^L(x,y,q)=i\sigma_0\dfrac{3y}{16x}\Bigg\{
2\Big[1+3\dfrac{(x+iy)^2}{q^2}+\dfrac{q^2}{4}\Big]-\hspace{4cm}$$$$-
\dfrac{1}{q^5}\Bigg[q^2-\Big(\dfrac{q^2}{2}-x-iy\Big)^2\Bigg]^2
\ln\dfrac{q^2/2-x-iy+q}{q^2/2-x-iy-q}-$$$$-
\dfrac{1}{q^5}\Bigg[q^2-\Big(\dfrac{q^2}{2}+x+iy\Big)^2\Bigg]^2
\ln\dfrac{q^2/2+x+iy+q}{q^2/2+x+iy-q}\Bigg\}.
\eqno{(5.8)}
$$

Subtracting from (5.8) gauge summand, for an integrated part
of conductivity $\hat\sigma_2$ we receive expression
$$
\hat\sigma_{2}^L(x,y,q)=i\sigma_0\dfrac{3y}{16x}\Bigg\{
2\Big[-\dfrac{5}{3}+3\dfrac{(x+iy)^2}{q^2}+\dfrac{q^2}{4}\Big]-\hspace{3cm}$$$$-
\dfrac{1}{q^5}\Bigg[q^2-\Big(\dfrac{q^2}{2}-x-iy\Big)^2\Bigg]^2
\ln\dfrac{q^2/2-x-iy+q}{q^2/2-x-iy-q}-$$$$-
\dfrac{1}{q^5}\Bigg[q^2-\Big(\dfrac{q^2}{2}+x+iy\Big)^2\Bigg]^2
\ln\dfrac{q^2/2+x+iy+q}{q^2/2+x+iy-q}\Bigg\}.
\eqno{(5.9)}
$$

Let's demonstrate on Figs. 1 and 2 graphic  comparison
of expressions (5.6) and (5.9).

From formulas (5.6) and (5.9), and Figs. 1 and 2 it's clear, that
these expressions differ not only analytically, but also numerically.
Hence, the  expression for conductivity from \cite {Dressel},
is incorrect.

Let's return to equality (5.2) and we will present it in the form
$$
{\sigma}_{tr}^{(1)}(x,y,q)=i\sigma_0\dfrac{y}{x}+{\sigma}_2(x,y,q),
\eqno{(5.10)}
$$
where ${\sigma}_2(x,y,q)$ is defined by the equality (5.6).

Let's rewrite the formula (5.10) by means of (5.6) in the explicit form:
$$
{\sigma}_{tr}^{(1)}(x,y,q)=
i\sigma_0\dfrac{3y}{8x}\Bigg[1+3\dfrac{z^2}{q^2}+
\dfrac{q^2}{4}+\dfrac{1}{2q^5}\Big[\Big(z^2-q^2+\dfrac{q^4}{4}\Big)^2+
z^2q^4\Big]\times
$$
$$
\times\ln\dfrac{z^2-(q-q^2/2)^2}{z^2-(q+q^2/2)}+
\dfrac{z}{q^3}\Big(z^2-q^2+\dfrac{q^4}{4}\Big)\ln\dfrac{(z-q)^2-q^4/4}
{(z+q)^2-q^4/4}\Bigg].
\eqno{(5.11)}
$$

Now the transverse Lindhard electric conductivity
is defined by expression (5.9).

The difference of conductivities (4.20) and (5.9) is equal to:
$$
\sigma_{tr}-\sigma_{tr}^{(1)}=\sigma_0\dfrac{3y^2}{4x}
\Bigg[\dfrac{2z}{q^2}+\dfrac{1}{q^3}(z^2-q^2)\ln\dfrac{z-q}{z+q}\Bigg].
$$

This equality shows, that at increase $q$ the difference
$\sigma_{tr}-\sigma_{tr}^{(1)}$ tends to zero.

Let's show comparison of the obtained expressions of conductivity.
For this purpose let's take advantage of equalities (4.20), (5.11) and
(4.8).
On following three plots curves of $1,2$ and $3$ answer accordingly
electric conductivity, constructed according to (4.20), (5.11) and
(4.8).

From Figs. 3 and 4, and also from Figs. 5 and 6 one can see,
that at small values
$q$ curves of $1$, answering (4.21), coincide with curves of $3$,
answering (4.8), and at large $q$ curves of $1$ coincide with
curves of $2$, answering to  Lindhard's expression (5.10).

From Figs. 7 and 8 it is clear, that at large values of
dimensionless frequency $x$ and at large values $q$ the curves
$1,2,3$ coincide  among themselves.

On Figs. 9 and 10 dependences of the real and imaginary parts
of transverse conductivity on dimensionless frequency $x$ at the various
values of parameter $q$ are presented. The curves $1,2,3$
correspond to values $q=0.1, 1,2$ accordingly.

\begin{center}
\bf 7. CONCLUSION
\end{center}

In the present work the correct formula for calculation of
transverse electric conductivity in the quantum collisinal
plasma  is deduced.
For this purpose the Wigner --- Vlasov --- Boltzmann kinetic
equation with collisional integral in the form of BGK--model
(Bhatnagar, Gross and Krook) in coordinate space is used.
The case of degenerate plasma is considered  separately.
Comparison with Lindhard's formula has been
realized.
\begin{figure}[bh]\center
\includegraphics[width=16.0cm, height=12.5cm]{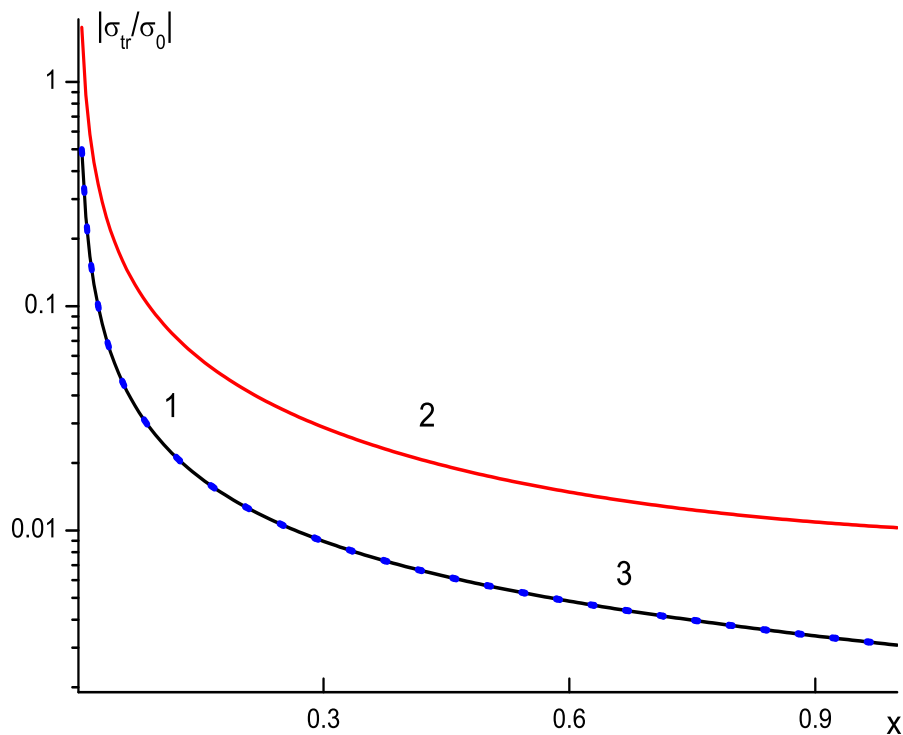}
\noindent\caption{The case: $q=2,y=0.01.$
Dependence $|\sigma_2/\sigma_0|$ on the dimensionless frequency
$x$.}\label{rateIII}
\end{figure}~
\begin{figure}[t]
\includegraphics[width=16.0cm, height=10.cm]{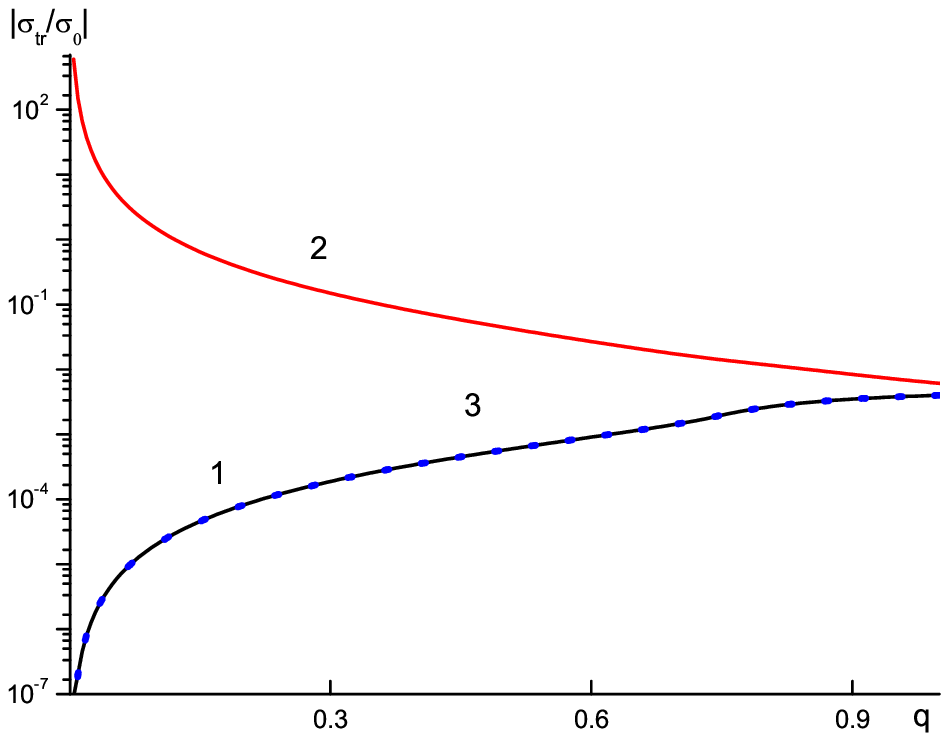}
\noindent\caption{The case: $x=1,y=0.01.$
Dependence $|\sigma_2/\sigma_0|$ on
the dimensionless wave number $q$.}
\includegraphics[width=16.0cm, height=10.cm]{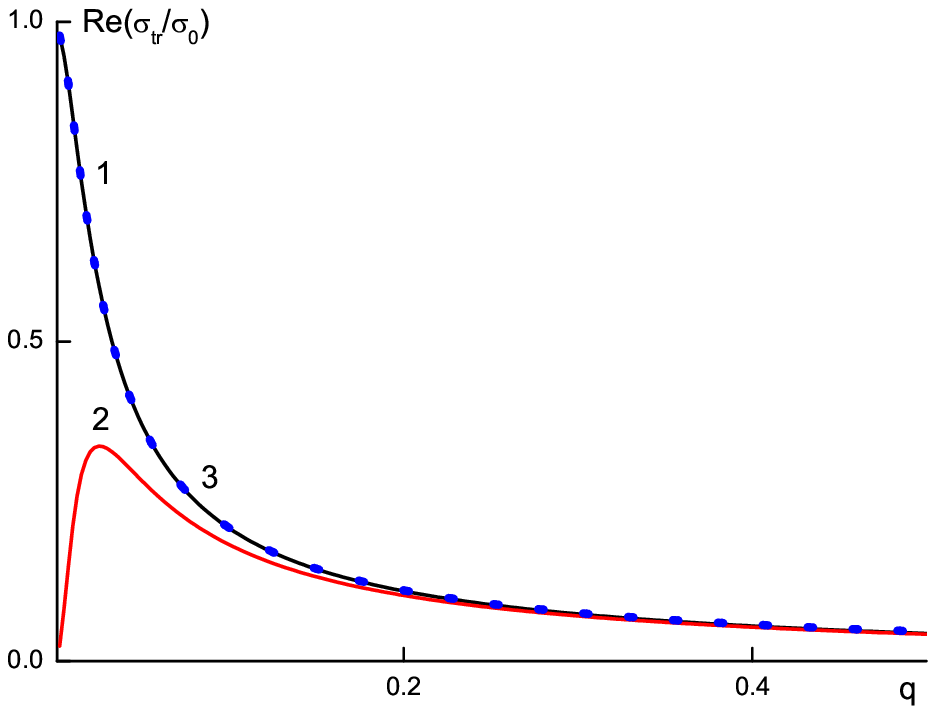}
\caption{The case: $x=0.001,y=0.01.$
Dependence  $\Re(\sigma_{tr}/\sigma_0)$ on the dimensionless wave
number $q$.}\label{rateIII}
\end{figure}

\begin{figure}[t]
\includegraphics[width=16.0cm, height=10cm]{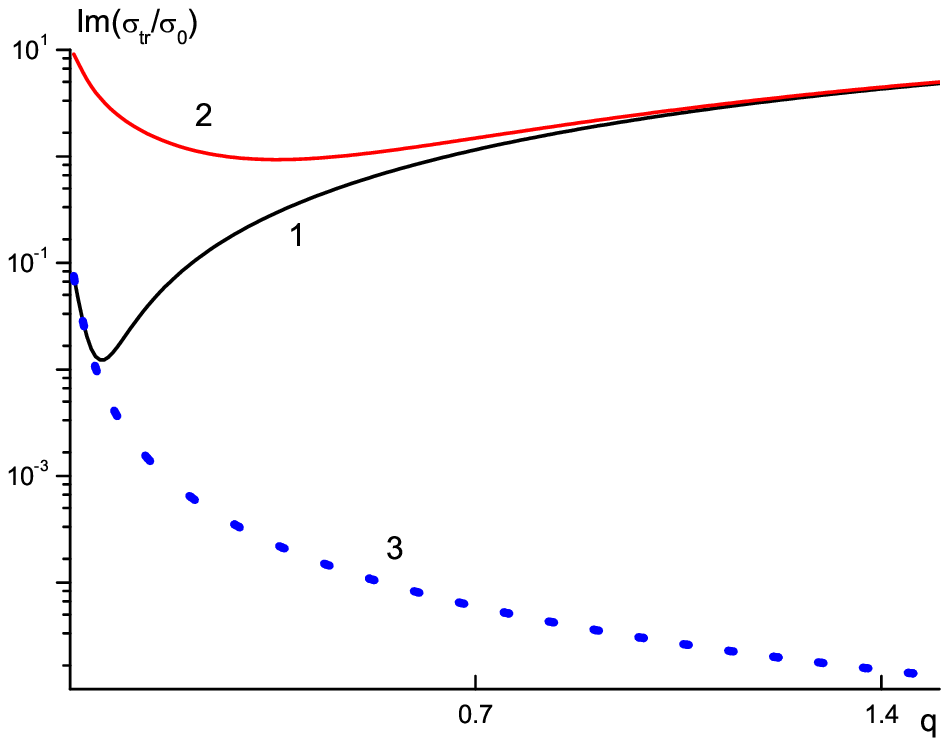}
\caption{The case $x=0.001,y=0.01.$
Dependence  $\Im(\sigma_{tr}/\sigma_0)$ on the dimensionless wave
number $q$.}
\includegraphics[width=16cm, height=10.cm]{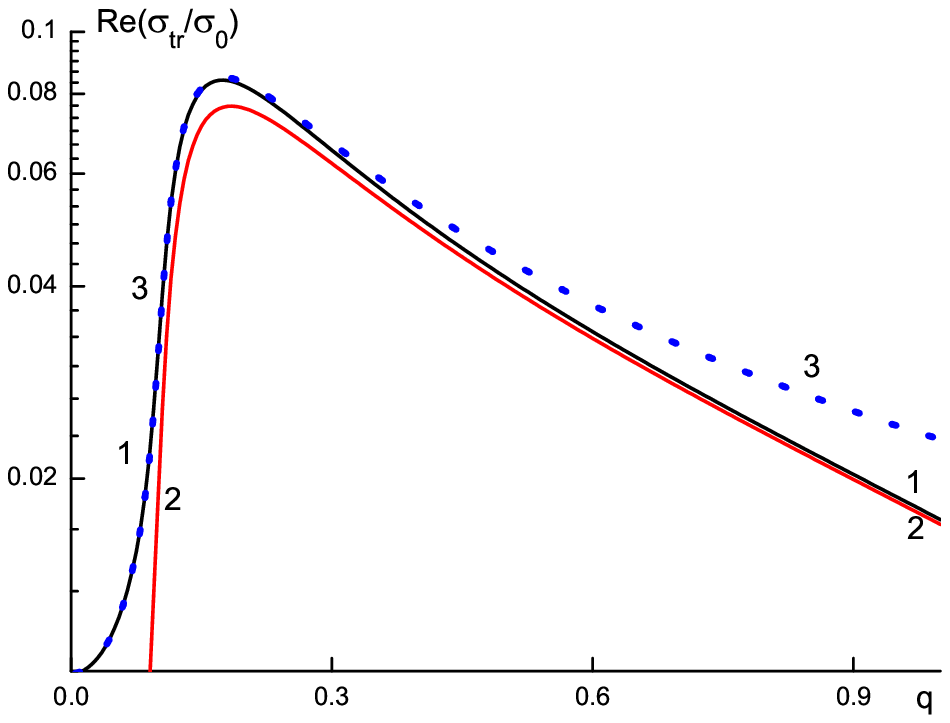}
\caption{The case: $x=0.1,y=0.01.$
Dependence  $\Re(\sigma_{tr}/\sigma_0)$ on the dimensionless wave
number $q$.}
\end{figure}

\begin{figure}
\includegraphics[width=16cm, height=10.cm]{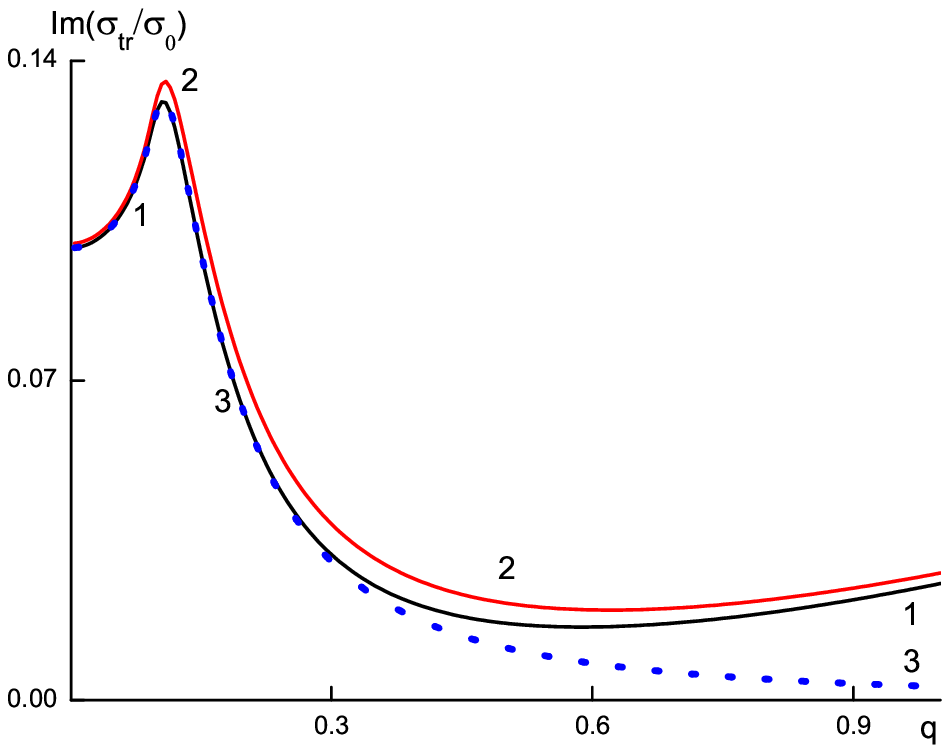}
\caption{The case: $x=0.1,y=0.01.$
Dependence  $\Im(\sigma_{tr}/\sigma_0)$ on the dimensionless wave
number $q$.}
\includegraphics[width=16cm, height=10.cm]{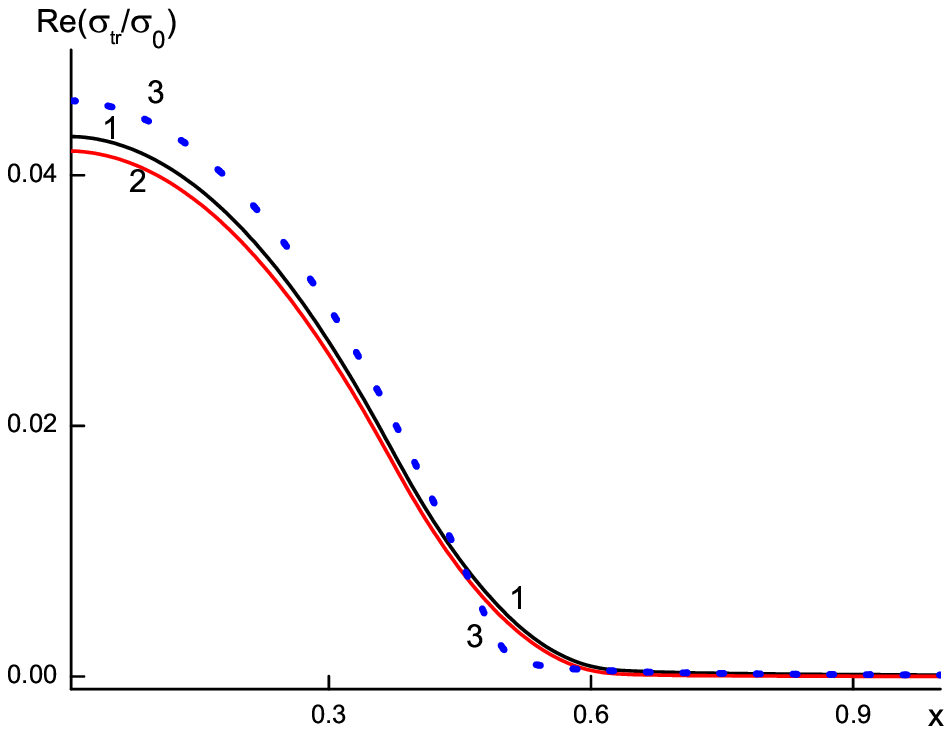}
\caption{The case: $y=0.01,q=0.5.$
Dependence  $\Re(\sigma_{tr}/\sigma_0)$ on dimensionless frequency $x$.}
\end{figure}

\begin{figure}[h]
\includegraphics[width=16cm, height=10.cm]{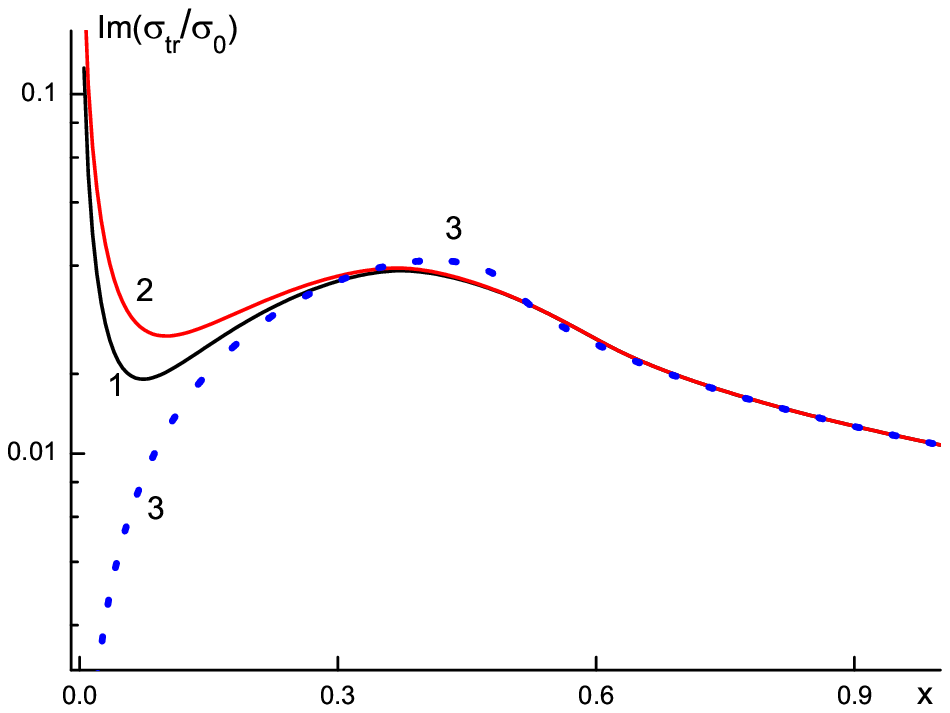}
\caption{The case $y=0.01,q=0.5.$
Dependence  $\Im(\sigma_{tr}/\sigma_0)$ on dimensionless frequency $x$.}
\includegraphics[width=16cm, height=10.cm]{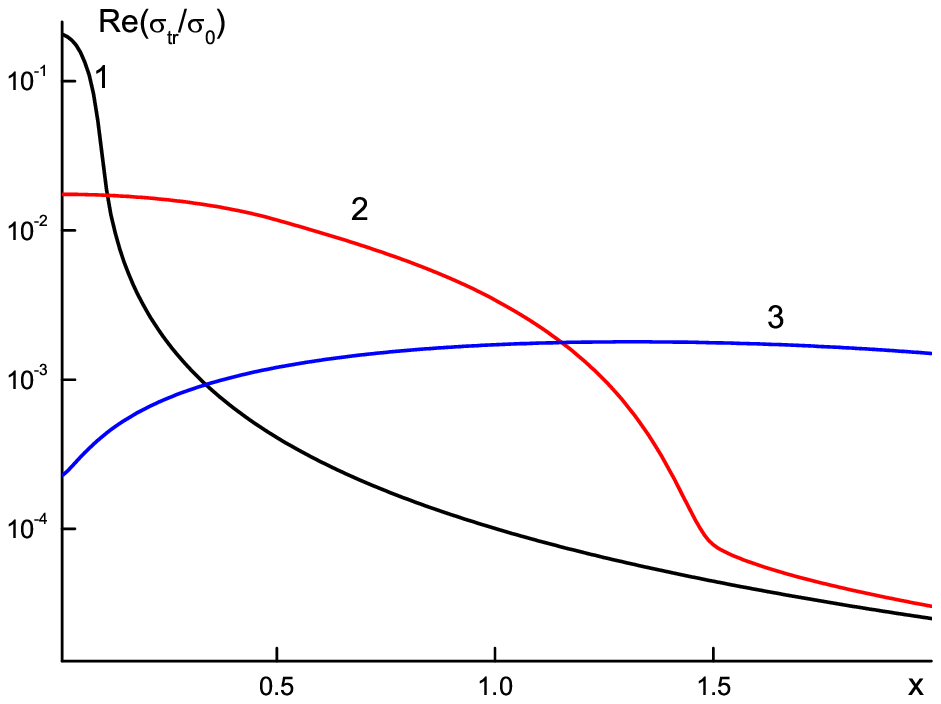}
\caption{The case: $y=0.01.$
Dependence  $\Re(\sigma_{tr}/\sigma_0)$ on dimensionless frequency $x$.}
\end{figure}

\begin{figure}[h]
\includegraphics[width=16cm, height=20.cm]{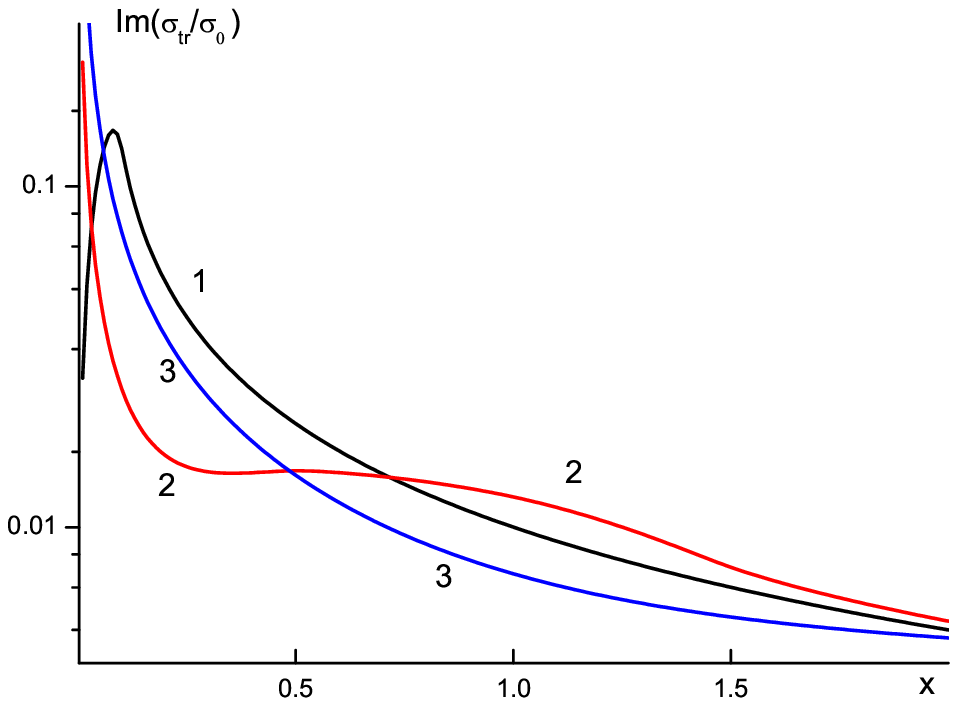}
\caption{The case: $y=0.01.$
Dependence   $\Im(\sigma_{tr}/\sigma_0)$ on dimensionless frequency $x$.}
\end{figure}

\end{document}